\documentclass[conference]{IEEEtran}

\def\BibTeX{{\rm B\kern-.05em{\sc i\kern-.025em b}\kern-.08em
    T\kern-.1667em\lower.7ex\hbox{E}\kern-.125emX}}

\usepackage{graphicx}
\usepackage{xcolor}
\usepackage{listings}
\usepackage[numbers]{natbib}
\definecolor{LightGray}{gray}{0.9}

\usepackage{stfloats}

\usepackage{caption}
\usepackage{subcaption}
\usepackage{url}

\usepackage{multicol}
\usepackage{fancyhdr}

\newcommand{\change}[1]{\textcolor{black}{#1}}

\newenvironment{code}{\captionsetup{type=listing}}{}

\definecolor{mGreen}{rgb}{0,0.6,0}
\definecolor{mGray}{rgb}{0.5,0.5,0.5}
\definecolor{mPurple}{rgb}{0.58,0,0.82}
\definecolor{backgroundColour}{rgb}{0.95,0.95,0.92}

\lstdefinestyle{CStyle}{
    backgroundcolor=\color{backgroundColour},   
    commentstyle=\color{mGreen},
    keywordstyle=\color{magenta},
    numberstyle=\tiny\color{mGray},
    stringstyle=\color{mPurple},
    basicstyle=\ttfamily,
    breakatwhitespace=false,         
    breaklines=true,                 
    captionpos=b,                    
    keepspaces=true,                 
    numbers=left,                    
    numbersep=5pt,                  
    showspaces=false,                
    showstringspaces=false,
    showtabs=false,                  
    tabsize=2,
    language=C
}

\title{Keyless Entry: Breaking and Entering\\ eMMC RPMB with EMFI}

\author{\IEEEauthorblockN{Aya Fukami}
\IEEEauthorblockA{\textit{Netherlands Forensic Institute}\\
\textit{University of Amsterdam}\\
}
\and
\IEEEauthorblockN{Richard Buurke}
\IEEEauthorblockA{\textit{Netherlands Forensic Institute} \\
}

}
\begin{document}
\maketitle

\setlength{\headheight}{18pt}
\maketitle
\thispagestyle{fancy}

\begin{abstract}
The Replay Protected Memory Block (RPMB) in modern storage 
systems provides a secure area where data integrity is ensured by authentication. This block is used in digital devices to store  pivotal information that must be safeguarded against modification by potential attackers. This paper targets the authentication scheme of the RPMB in three different eMMCs from a major manufacturer.
A glitch was injected by sending an electromagnetic pulse to the target chip. RPMB authentication was successfully glitched and the information stored in two target eMMCs was overwritten with arbitrary data, without affecting the integrity of other data. 
\end{abstract}


\section{Introduction}
\label{section:introduction}

In modern smart devices, commonly used storage systems often implement a Replay Protected Memory Block (RPMB). The RPMB is a hardware partition in modern storage devices such as Embedded Multi Media Card (eMMC), Universal Flash Storage (UFS), and Non-volatile Memory express (NVMe). 

\begin{figure}[!ht]
\centering
    \includegraphics[width = .95\linewidth]{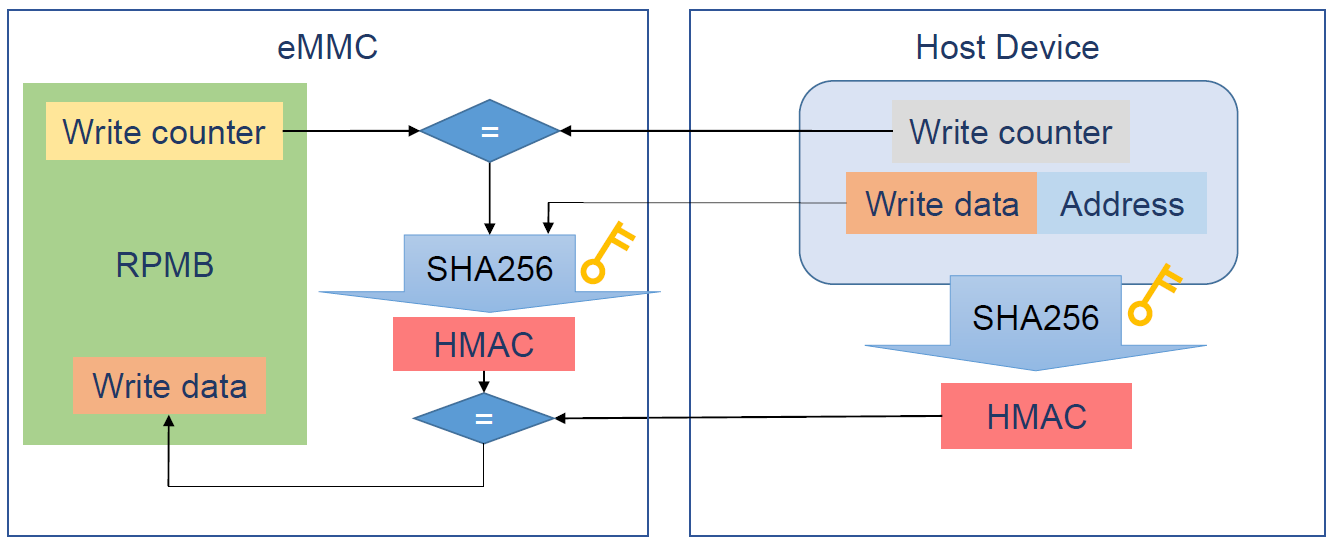}
        \vspace{1pt}
    \caption{RPMB write sequence block diagram}
    \label{fig:rpmb-scheme}
\end{figure}

Writing data to the RPMB requires authentication using a cryptographic hash function, namely a keyed-hash message authentication code (HMAC) using SHA256. The HMAC is calculated over the data frame, excluding some fields such as the HMAC itself. The key used for the calculation is programmed into the storage device only once. In modern embedded devices, because RPMB authentication relies solely on the confidentiality of the pre-shared key, a secure component such as a Trusted Execution Environment (TEE) takes ownership of the RPMB \cite{qc_rpmb}. Figure \ref{fig:rpmb-scheme} shows the basic concept of accessing the RPMB. 

\change{By issuing a RPMB read request command, anyone can read the content of the RPMB data.} Therefore the RPMB is not suitable for storing confidential information. Rather, the RPMB is commonly used to store information that is immutable for a normal user, for example:

\begin{itemize}
    \item An anti-rollback counter / version information \cite{DFRWS2024,giese2021amazon}
    \item Cryptographic public keys \cite{ConnectCore} 
    \item Bootloader lock state \cite{uboot_rpmb}
\end{itemize}

As an example, an anti-rollback counter stored in the RPMB can be used as part of the key derivation process. When the device is wiped, because of a factory reset or by exceeding the maximum number of allowed password attempts, the anti-rollback counter in the RPMB is incremented. In this scenario, even when restoring the entire contents of flash user data, key derivation will fail and the encrypted user data cannot be decrypted \cite{DFRWS2024}. 


Since authentication is performed by a cryptographic hash function using a pre-shared key, integrity of the information stored in the RPMB is guaranteed as long as the secrecy of the key is maintained.
This research tries to break this authentication scheme, enabling an attacker to overwrite data in the RPMB without knowledge of the pre-shared key. \change{We targeted the RPMB in eMMCs due to their availability,} and usage in a wide variety of embedded products, such as smartphones \cite{DFRWS2024}, IoT devices \cite{giese2021amazon} and automotive systems \cite{NXP_automotive}. 

Several techniques for attacking RPMB authentication exist, and will be covered more extensively in Section \ref{section:related-work}. For this experiment fault injection (FI) was applied, which is an umbrella term for a collection of techniques that aims to introduce faults, or glitches, into a device leading to unintended behavior. This can be achieved through software \cite{clkscrew,rowhammer} or hardware methods \cite{SHEPHERD2021102471, sass2023oops, Pengfei2019}. Commonly used hardware methods are:

\begin{itemize}
    \item Shorting the power supply (crowbar glitching, or voltage fault injection (VFI))
    \item Introducing electromagnetic pulses (EMFI)
    \item Illumination with a laser beam (LFI)
    \item Changing the clock signal 
\end{itemize}

Fault injection can be used to circumvent security checks in software whilst the program itself does not contain any (known) vulnerabilities. Unintended behavior that can aid the attacker is also referenced as a fault primitive, such as skipping instructions or corrupting CPU register values.  

By applying EMFI on an eMMC device we were able to write arbitrary data to the RPMB, without knowledge of the pre-shared key and thereby breaking the RPMB authentication scheme which compromises the integrity of all stored data. We then applied the same method to a different model eMMC device which was also successfully compromised.

This research makes the following contributions:

\begin{itemize}
    \item \change{We show how EMFI can be applied on an eMMC device to enable writing of arbitrary data to the RPMB, using commercially available tools.}
    \item A detailed description on how to characterize the susceptibility \change{to} EMFI, on Samsung eMMC controllers, through arbitrary code execution is provided.
    \item Advantages and disadvantages of applying EMFI, compared to other methods for breaking RPMB authentication are outlined.
\end{itemize}


The structure of the rest of this paper is as follows. \change{In Section \ref{section:threat_model}, we discuss the threat model related to our research.} Section \ref{section:related-work} discusses related work to this research. The targets and physical FI setup are described in Section \ref{section:experiment-setup}. \change{Section \ref{section:characterize} outlines how the firmware of our target devices was obtained, and reversed engineered.} This section also describes how this information was used to run our own fault observer code on the target, to characterize the susceptibility to EMFI. The process for successfully applying EMFI on two of the targets, is detailed in Section \ref{section:experiment}. 
The impact of this attack is discussed in section \ref{section:discussion} before concluding in section \ref{section:conclusion}.

\begin{figure*}[bp]
\centering
    \includegraphics[width=0.85\linewidth]{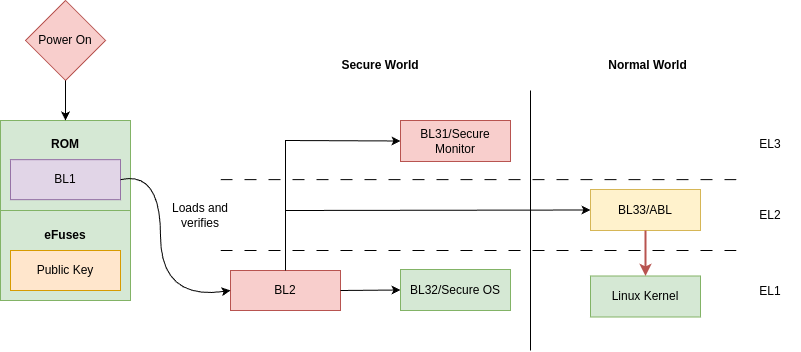}  
    \vspace{1pt}
    \caption{ARM Trusted Firmware-A secure boot overview}
    \label{fig:secure-boot}
\end{figure*}

\section{Threat Model}
\label{section:threat_model}
\change{Because the RPMB is a type of authenticated storage, it is often used for storing non-confidential information that requires integrity protection. Examples include version information used in anti-rollback mechanisms \cite{DFRWS2024, uboot_rpmb, nxp_imx_security_guide}, bootloader lock state \cite{uboot_rpmb,nxp_imx_security_guide} and asymmetric public keys \cite{ConnectCore,nxp_imx_security_guide}.}
\change{Modification of this type of information by an unauthorized attacker can result in an increased attack surface, or in some cases, a fully compromised system.} 
\change{Fukami, et al. (2024) \cite{DFRWS2024} showed that it is possible to decrement an anti-rollback counter stored in the RPMB to return a wiped Android device to a non-wiped state. Anti-rollback mechanisms are also often used to prevent software components from being downgraded to a vulnerable state \cite{AndroidRollbackProtection, TrustIssues}.} 

\change{Modern Android systems are commonly based on the ARM Trusted Firmware-A reference design, which implements a chain-of-trust (CoT) for the boot process \cite{ArmTFACoT}. Figure \ref{fig:secure-boot} shows a simplified representation of a typical secure boot implementation.}
\change{Devices based on the ARM Cortex-A architecture feature a system-on-chip (SoC) which can switch between multiple contexts. The normal operating system (OS) runs in the so-called "normal world" while a dedicated second operating system is executed in the "secure world". This separation provides an additional layer of security since the OS running in normal world cannot access memory and resources used by the secure world.}

\change{After powering on the device, the first boot stage, embedded in the ROM of the SoC, is executed. The secondary boot stage is loaded from flash and verified using a public key stored in eFuses \cite{nakamoto2016secure}. Each subsequent boot image is verified before being executed.}
\change{The Android Bootloader (ABL) implements Android Verified Boot (AVB), and is responsible for starting the Android operating system. If the bootloader is locked, a public key verifies the integrity of the boot image, and hashtrees for the system and vendor partitions \cite{AVB2}.}
\change{Some vendors, such as NXP \cite{nxp_imx_security_guide}, recommend storing the bootloader state and AVB key in the RPMB area of an eMMC. By manipulating this information an attacker can unlock the bootloader or modify and resign the Android operating system, resulting in code execution with EL1 privileges in the normal world.}

\begin{table*}[tp]
\caption{Target chip properties}
\label{table:device_id}
\vspace{-2mm}
\centering
\begin{tabular}{|c|c|c|c|}
    \hline
    Target number & 1 & 2 & 3 \\
    \hline
    Product name & KLMAG2GE4A & KLMBG2JETD &  KLM8G1WEMB\\
    \hline
    Manufacturer ID & 0x15 & 0x15 & 0x15 \\
    \hline
    eMMC version & 4.41 & 5.1 & 5.0 \\
    \hline
    Part name & MAG2GA (1.2) & BJTD4R (5.6)  & 8WMB3R (0.0) \\
    \hline
    Part no. & 0xc23 (Cortex-M3) & 0xc27 (Cortex-M7) & 0xc23 (Cortex-M3) \\
    \hline
    Architecture & 0x0F (ARMV7-M) & 0x0F (ARMV7-M) & 0x0F (ARMV7-M) \\
    \hline
    Variant & 0x02 & 0x01 & 0x02 \\
    \hline
    Revision & 0x00 & 0x01 & 0x00 \\
    \hline
    Controller name & VHX0 & VCT0 & VPX0 \\
    \hline
    MPU enabled & No & No & No \\
    \hline
    VTOR & 0x40000 & 0x60000000 & 0x40000\\
    \hline
\end{tabular}

\end{table*}
\section{Related Work}
\label{section:related-work}
Recovering the secret key has been the main focus of security research related to HMAC authentication. \citeauthor{kitae2013} showed that they could successfully recover the key by injecting faults and reducing the number of computational rounds to HMAC \citep{kitae2013}. \citeauthor{yaacov2021} published their research on a side channel attack against HMAC. The authors successfully simulated recovery of the key through a template attack \cite{yaacov2021}. Our research focuses on the application of HMAC in a real-world device, where the HMAC authentication itself needs to be skipped. 

Western Digital published a white paper on vulnerabilities in the eMMC RPMB \cite{western2}, and suggested that by performing a ``man-in-the-middle" attack, one can trick the host system to believe that writing the data was successful, whilst in the background different data is being written. Their attack is effective if targeting a ``live'' device, where an attacker can monitor the command issued by the host. In our scenario, arbitrary data is written to the RPMB without relying on the behavior of the host device.

Fault injection attacks against mobile or IoT devices have been widely performed by multiple researchers, as reported by \citeauthor {electronics11132023} and \citeauthor{SHEPHERD2021102471} \cite{electronics11132023,SHEPHERD2021102471}. In their research, SoCs are the main target to introduce glitches in order to skip the cryptographic authentication and gain root privileges on the target device.

\change{Multiple publications discuss attacking ARM Cortex-M microcontrollers. \citeauthor{bozzato2019shaping} showed that it is possible to improve the effectiveness of voltage glitching attacks by injecting an arbitrary waveform instead of shorting the power supply \cite{bozzato2019shaping}. They validated their approach on multiple microcontrollers including an STM32F103 (Cortex-M3) and STM32F373 (Cortex-M4).}
\change{\citeauthor{ruminot2023novel} focused on improving voltage glitching attacks by reducing the voltage supplied to the target just outside of normal operating parameters \cite{ruminot2023novel}.} 
\change{\citeauthor{werner2023end}  created a tool that is able to find fault injection parameters in an automated manner, by combining characterization data with a simulated model \cite{werner2023end}. The authors applied their approach to a Cortex-M4 microcontroller, using a dual-laser fault injection setup.}

\change{Although the aforementioned research targets similar devices as in our research, they do not use EMFI. Instead, they opt for VFI or LFI. Furthermore, research by \citeauthor{ruminot2023novel}  \cite{ruminot2023novel} and \citeauthor{werner2023end} \cite{werner2023end} did not circumvent any real-life security mechanisms.}


\change{To the best of our knowledge, this is the first work to perform EMFI against eMMC RPMB authentication. The hardware designs of the targets are proprietary and unknown. We also did not have access to the source code, or reference implementations, of the firmware.}

\section{Experiment Setup}
\label{section:experiment-setup}
\subsection{Electromagnetic Fault Injection}
\change{We chose to perform EMFI for our attack scenario since it does not require any additional hardware modifications and provides localized injection of faults. LFI requires thinning of the package to expose the internal transistors, we therefore did not consider this method. The external clock signal might be susceptible to clock glitching, however we assumed that it was not directly connected to the application processor (AP) embedded in the controller.
Since an eMMC device is powered from an external source, the core voltage of the controller (Vddi) and memory peripherals (Vcc) can be trivially manipulated. Therefore it might also be possible to apply voltage glitching to these devices.}

\subsection{Target Selection}
\label{sec:target_selection}
The goal of this research is to bypass the cryptographic authentication of the RPMB in an eMMC. Prior to the actual attack, we decided to run a profiling process in order to identify the location of the chip where it is most susceptible to the EMFI attack. For this purpose, we selected eMMCs from a single manufacturer where we have access to their firmware.
We selected 3 eMMC devices shown in Table \ref{table:device_id} on availability in our forensic lab. 
We refer to these devices as Target 1, 2, and 3, respectively, for the rest of this paper.
Prior research \change{suggests that our target devices} contain a Cortex-M micro controller, and that the firmware of these devices can be read using vendor-specific commands \cite{Avraham_2018}. 
In 2018, \citeauthor{Avraham_2018} demonstrated that it was possible to read/write memory regions of \change{an eMMC controller}, embedded in a \change{mobile phone}, using proprietary vendor commands \cite{Avraham_2018}. The author subsequently released the proof-of-concept code on his public repository \cite{oranav2024Jan}. 
\change{We observed that these vendor commands are still implemented in an eMMC that is currently in mass-production. Therefore we assume that they are still applicable to a wide range of devices.}


\change{The firmware of each target device was extracted by using an EasyJTAG flasher box \cite{easyjtag}, which supports the required vendor-specific commands. }
\subsection{Glitching Setup}
\label{sec:glitchingsetup}
\begin{figure}[!ht]
\centering
    \includegraphics[width = .95\linewidth]{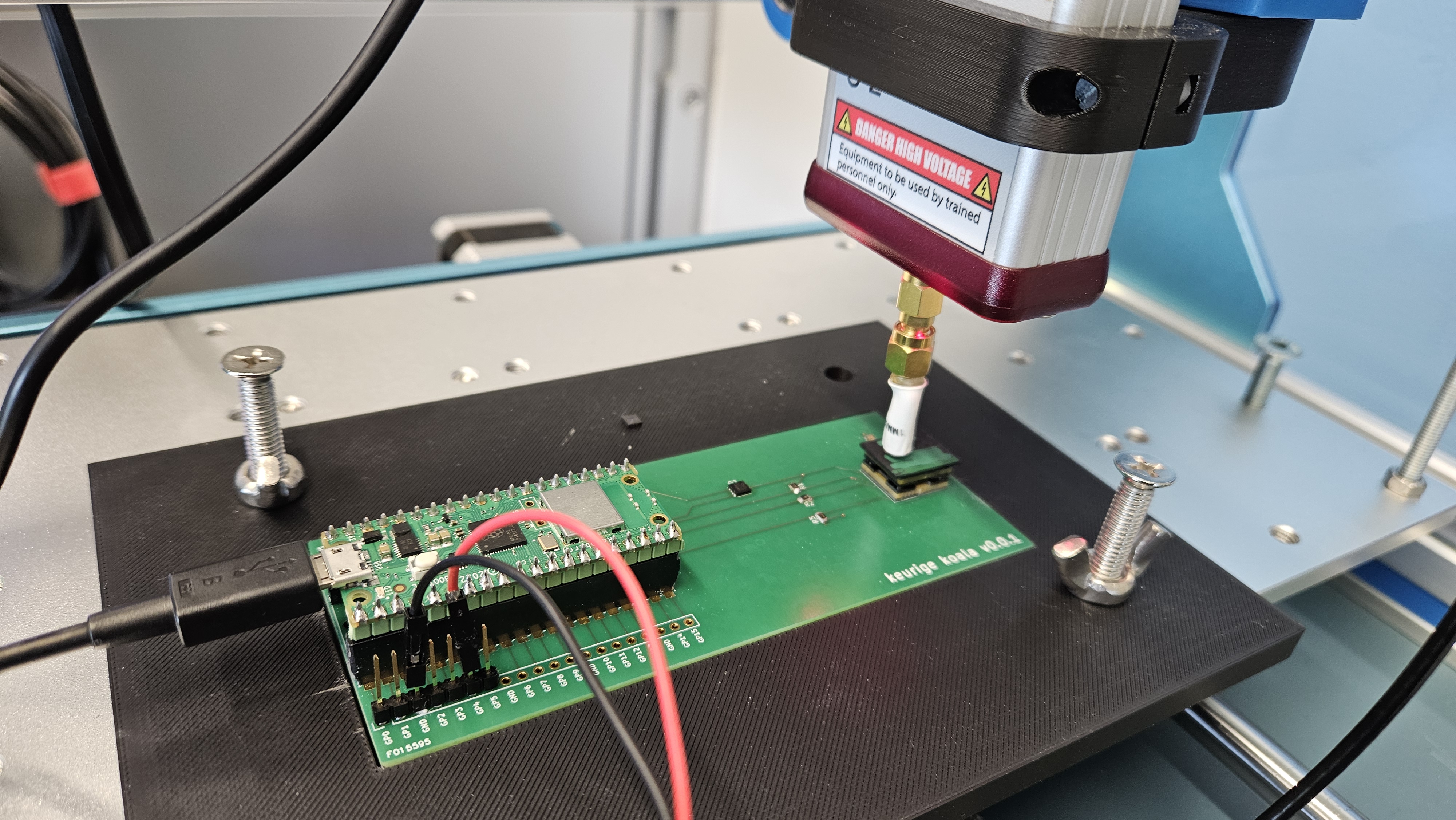}
    \vspace{-5pt}
    \caption{Target chip mounted to a custom PCB, using an adapter, while applying an EM pulse using a ChipSHOUTER (CW520)}
    \label{fig:fi-setup}
\end{figure}

\change{Each target eMMC was mounted on a custom breakout adapter in order to avoid the need for reballing and resoldering. The socket for the adapter is soldered onto a custom-made printed circuit board (PCB), which is mounted onto the XYZ table  (Genmitsu 3018 PROver V2) with a custom-made fixture for precise positioning and repeatability.}
Electromagnetic pulses were injected by a NewAE ChipSHOUTER (CW520) \citep{newae}, using a one millimeter clockwise winded probe. Figure \ref{fig:fi-setup} shows an overview of the setup.
\change{Communication with the target eMMC was implemented using a programmable I/O (PIO) based state machine, running on a Raspberry Pi Pico.} \change{The ChipSHOUTER is also triggered by the state machine program, using a fixed delay after sending the eMMC command. The rest of the fault injection process was orchestrated by software running on a Raspberry Pi 4.}

A Tektronix DPO7354C oscilloscope was used to monitor eMMC communication and measure EM emissions from the application processor (AP). EM measurements were used to determine the timing of eMMC operations, as described in Section \ref{section:glitching_setup}. However any four channel scope with around 200MHz bandwidth should suffice, such as a PicoScope 3000 Series.

Whilst industry level FI equipment is available, cost effective off-the-shelf products were used as much as possible for this attack setup.


\

\section{Characterization}
\label{section:characterize}
\change{Given that EMFI enables us to localize the attack, we first needed to identify the most vulnerable location of the target eMMC. We therefore acquired and analyzed the firmware of each target device, in order to implement a simple fault observer program that could run on the target device. The fault observer enabled us to monitor the effect of the EM pulse at different locations.}

\subsection{Device Identification}
We extended the \textit{mmc-utils} program \cite{mhei2024Jan} to include the vendor-specific commands discussed in \change{Section \ref{sec:target_selection}}. 
\change{We assumed that the flash controller} was based on the ARM architecture, and so our version of \textit{mmc-utils} was used to read the System Control Block (SCB), at offset 0xE000ED00, and the Memory Protection Unit (MPU) configuration, at offset 0xE000ED90. Table \ref{table:device_id} combines information from the SCB, MPU configuration and Device identification (CID) register.

\change{First, we noticed} that the MPU is disabled for all chips. According to the \textit{ARMv7-M Architecture Reference Manual} \cite{Arm2021Dec}, the default system memory map is used when the MPU is disabled. By default, the \textit{Code}, \textit{SRAM} and \textit{RAM} memory segments are mapped readable, writeable and executable.
The vector table offset register (VTOR) points to the main vector table of the device. The second entry in this table holds the address of the reset handler, which is the entry point for our ROM code.

\subsubsection*{Arbitrary Code Execution}
The code section also holds the vector table for standard eMMC commands and can therefore be overwritten using vendor-specific commands. For Target 2 this table was located in the RAM segment, but the same principles apply. To gain arbitrary code execution, the payload was first written to an unused memory region, and the entry for CMD8 (SEND\_EXT\_CSD) in the vector table was updated with the address of our routine.

\begin{code}
\begin{lstlisting}[style=CStyle, caption=Fault observer implementation, label=src:fault-observer]
void fault_observer(void) {
  uint32_t total_iterations;
  uint32_t value;
  uint32_t j;
  uint32_t i;
  extcsd *ext_csd;
  
  ext_csd = PTR_EXT_CSD;
  total_iterations = 0;
  value = 0;
  j = 0;
  do {
    j = j + 1;
    i = 0;
    do {
      value = value + 7;
      i = i + 1;
    } while ((int)i < 62500);
    total_iterations = total_iterations + i;
  } while ((int)j < 4);
  ext_csd->total_iterations = total_iterations;
  ext_csd->value = value;
  (*(code *)CMD8)();
  return;
}
\end{lstlisting}

\vspace{3pt}%
\end{code}

\change{Our fault observer implementation was directly written in assembly. Listing \ref{src:fault-observer} shows the decompilation of this code for readability.} It consists of a nested for loop that increments an unsigned integer value for every iteration. The total number of iterations and incremented value are written to the beginning of the extended CSD register, which according to the current JEDEC standard \cite{eMMC5ElecStandard} is unused. Finally the original CMD8 routine is executed, returning the contents of the extended CSD register. By checking the stored values, it is possible to determine if the controller was affected by the EM pulse. This approach worked for all targets. 

\subsection{Profiling}
\label{section:profiling}



\begin{figure*}[ht!]
    \begin{subfigure}{0.31\linewidth}
    \includegraphics[width=0.31\linewidth, trim={0, 150, 440, 0}]{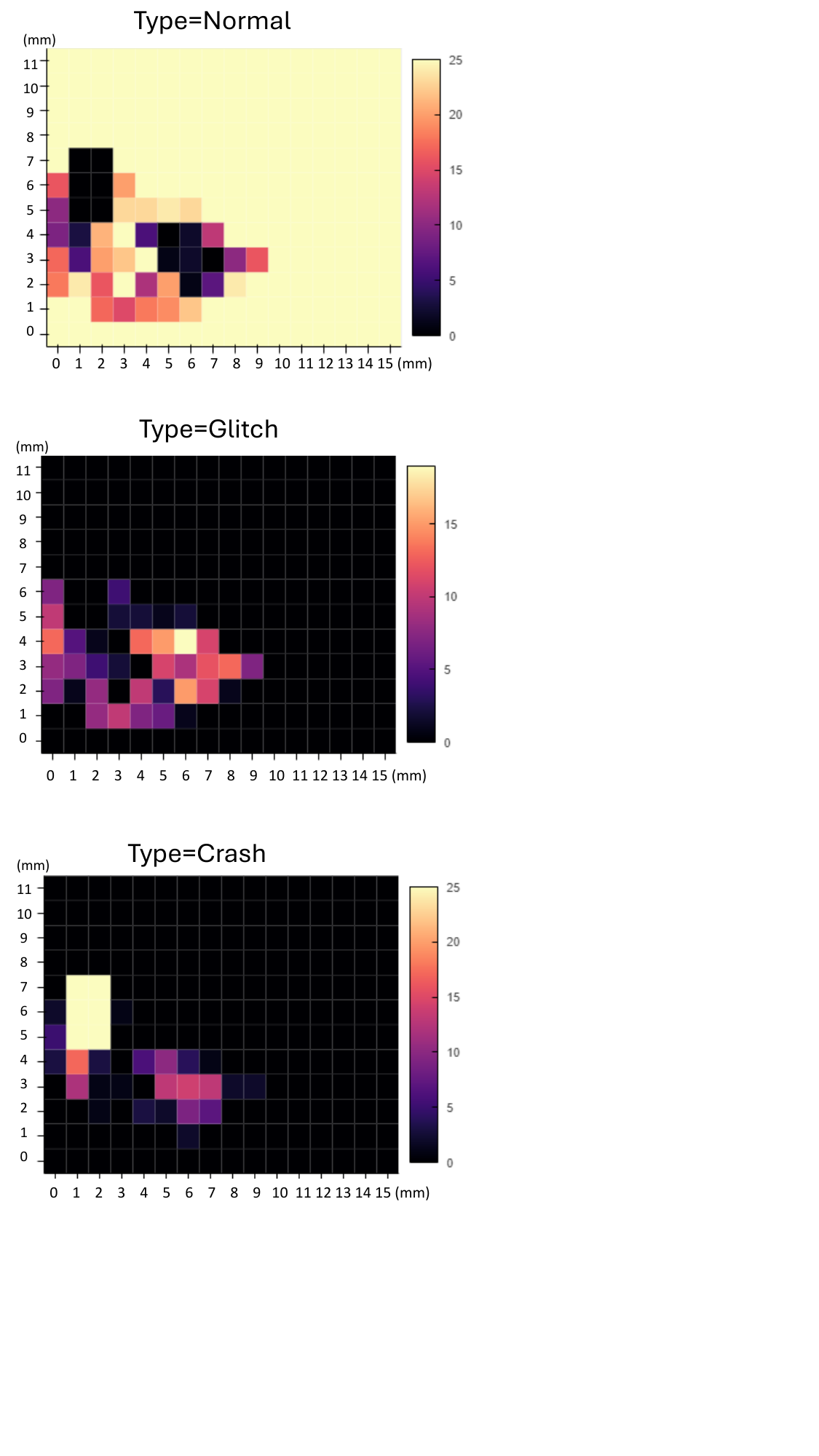}
    \caption{Profiling results: Target 1}
    \label{fig:heatmap-MAG2}
    \end{subfigure}
    \hfill%
    \begin{subfigure}{0.31\linewidth}
    \includegraphics[width=0.31\linewidth, trim={0, 150, 440, 0}]{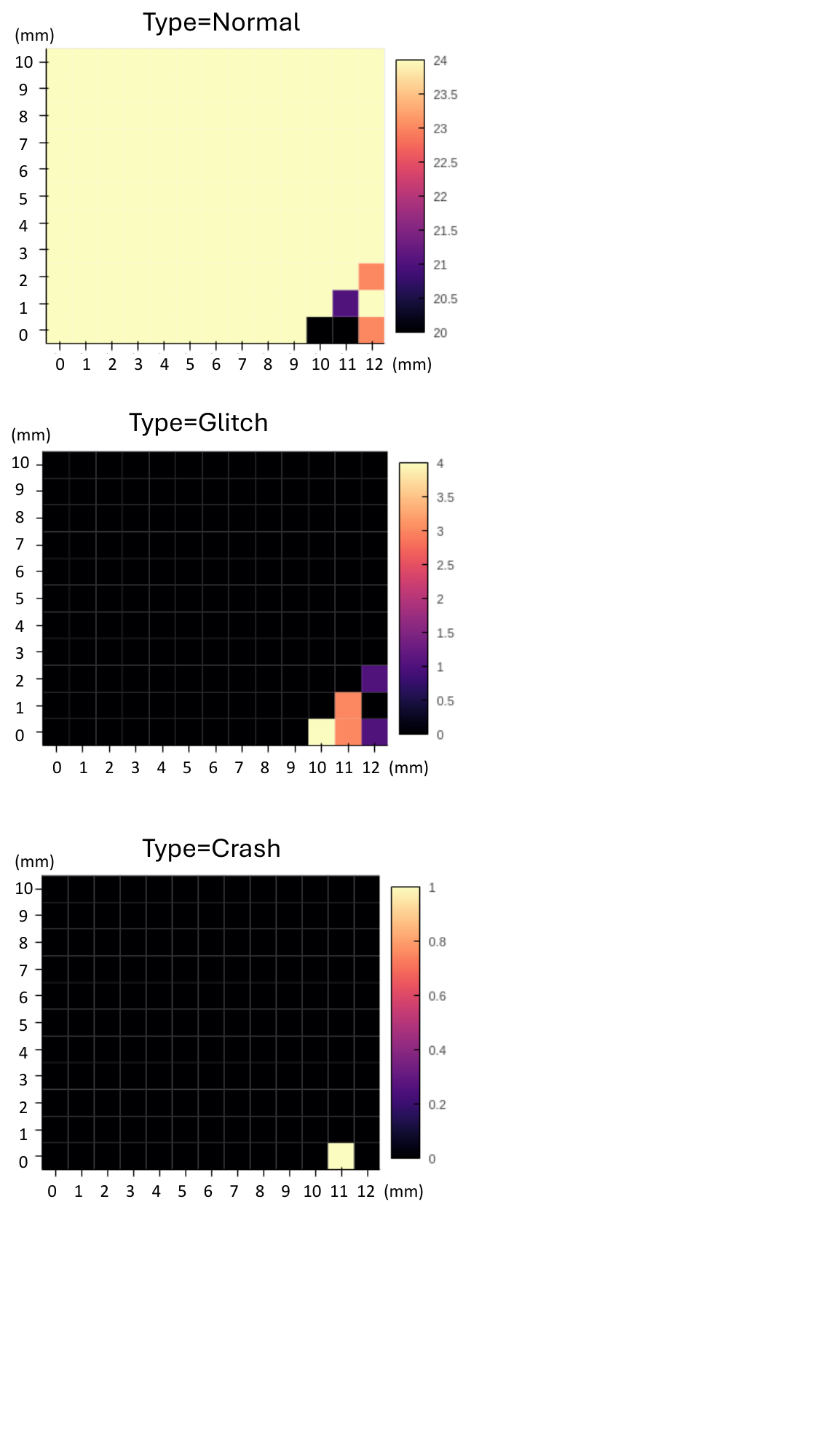}
    \caption{Profiling results: Target 2}
    \label{fig:heatmap-JET}
    \end{subfigure} 
    \begin{subfigure}{0.31\linewidth}
    \includegraphics[width=0.31\linewidth, trim={0, 150, 440, 0}, scale=0.9]{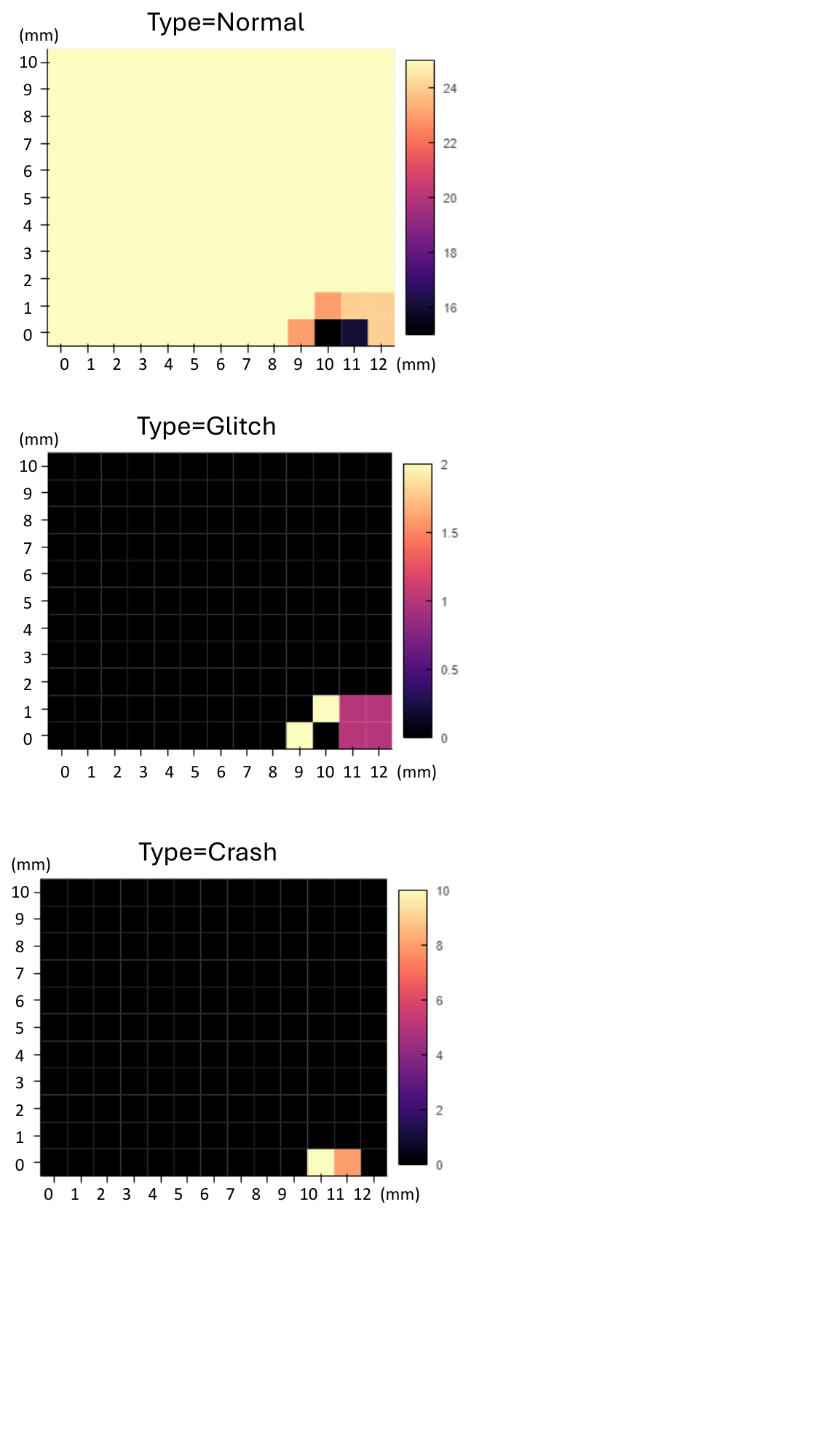}
    \caption{Profiling results: Target 3}
    \label{fig:heatmap-WEMB}
    \end{subfigure} 
    \caption{Profiling results for each chip using the fault observer. The lighter the color, the more susceptible the chip is for the categorized result at that location.}
    \label{fig:heatmap}
\end{figure*}

\begin{figure*}[h]
    \begin{subfigure}{0.33\linewidth}
    \includegraphics[width = 0.3\linewidth, trim={0, 80, 600, 0}]{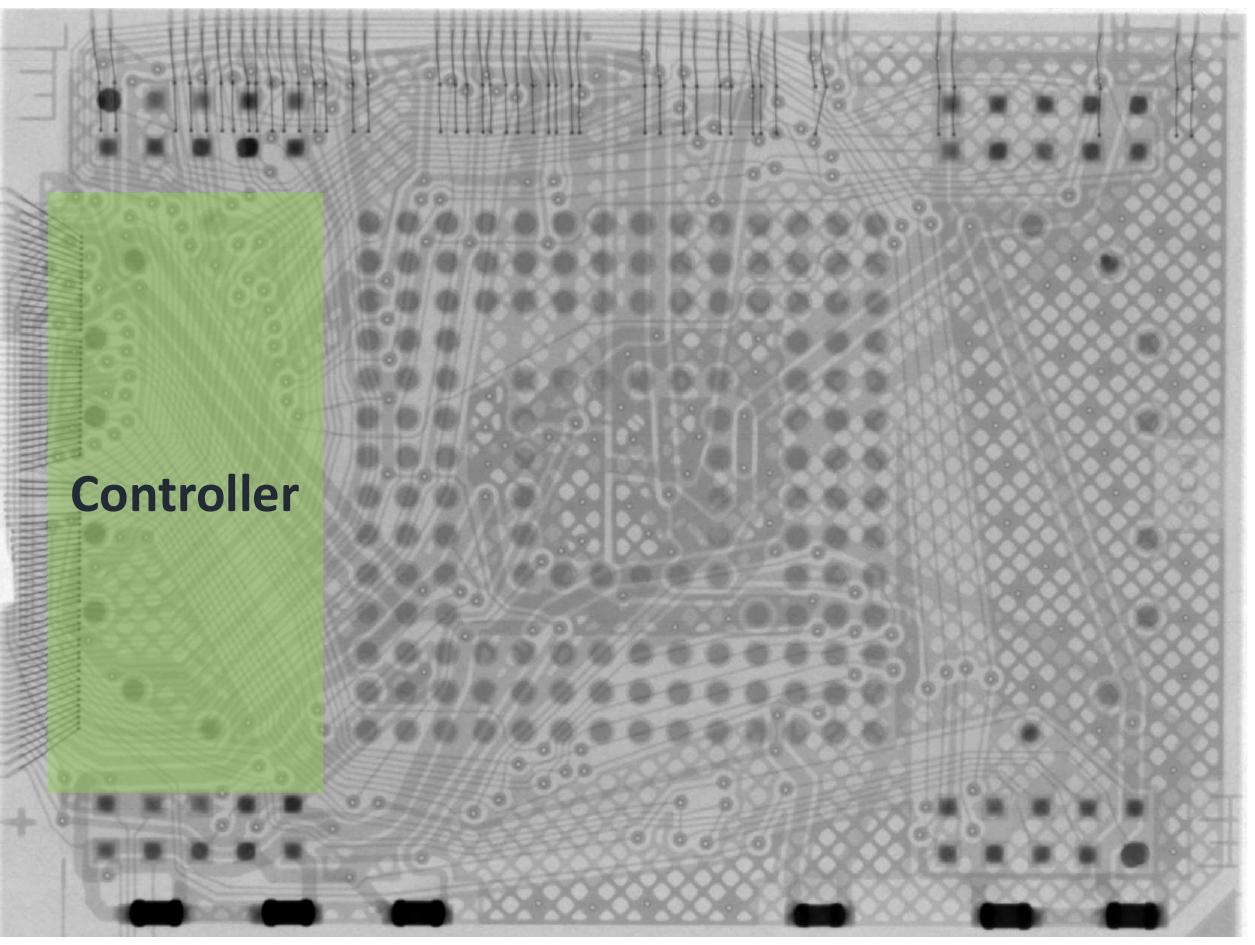}
    \caption{Target 1}
    \label{fig:xray1}
    \end{subfigure}%
    \centering
    \begin{subfigure}{0.33\linewidth}
    \includegraphics[width = 0.3\linewidth, trim={0, 80, 630, 0}]{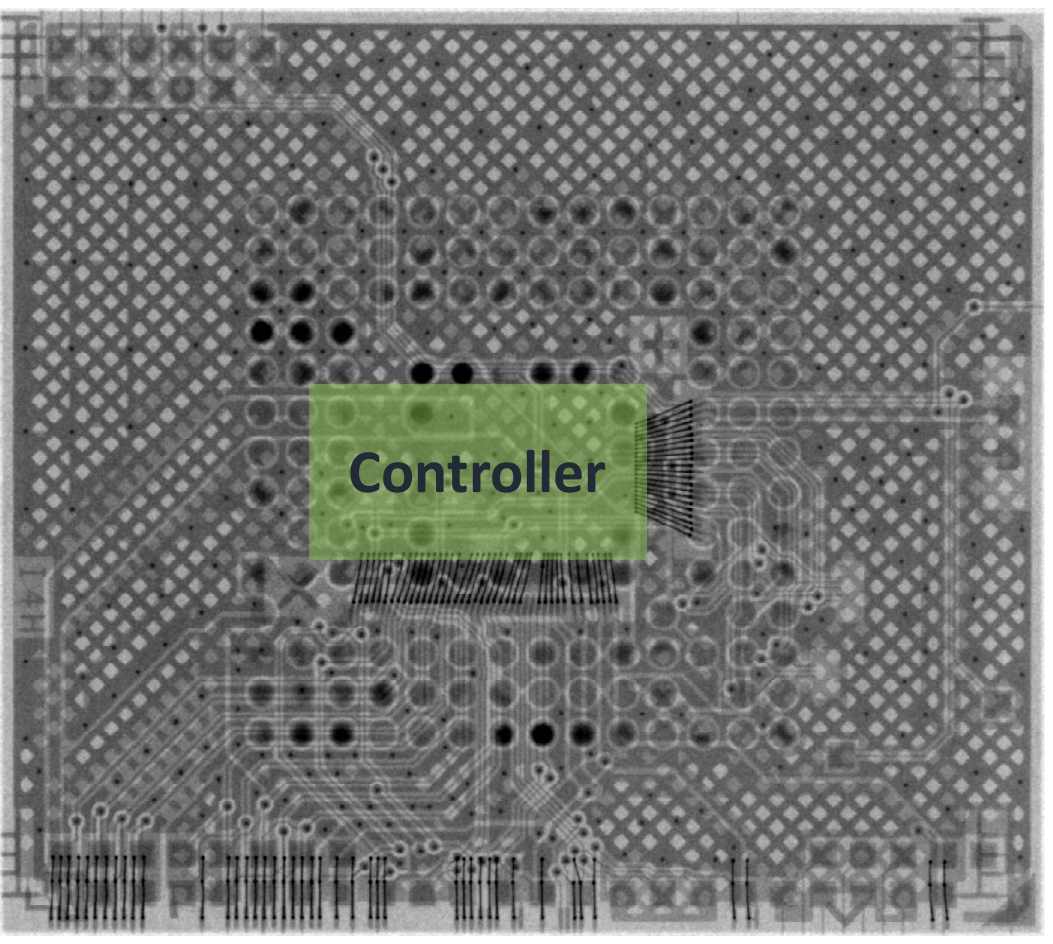}
    \caption{Target 2}
    \label{fig:xray2}
    \end{subfigure}
    \centering
    \begin{subfigure}{0.33\linewidth}
    \includegraphics[width = 0.3\linewidth, trim={0, 80, 630, 0}]{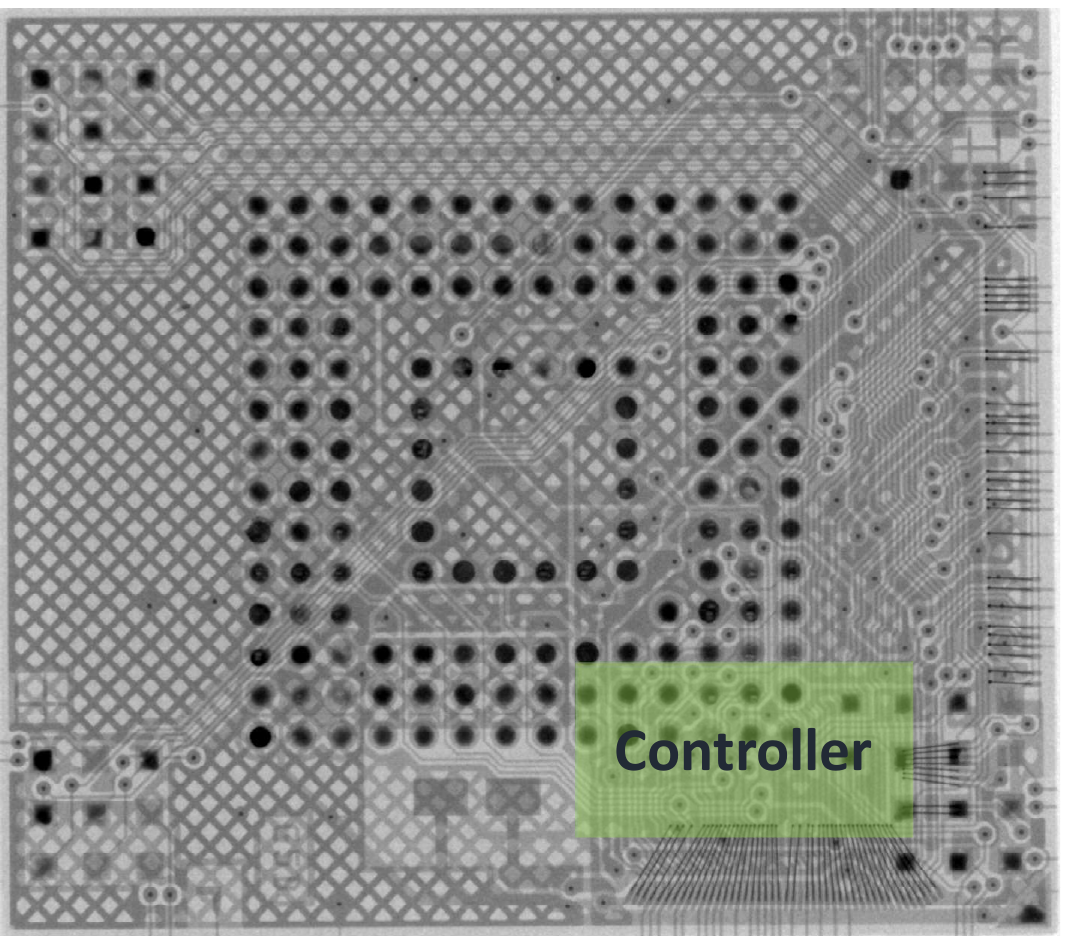}
    \caption{Target 3}
    \end{subfigure}
    \vspace{-8pt}
    \caption{X-ray inspection of target chips, highlighting the physical location of the controller}
    \label{fig:Target-xray}
\end{figure*}

By introducing EM pulses at different positions of the chip while the fault observer code was running on the target, \change{we determined which areas of the device were the most affected.}
Based on the return values of the fault observer, \change{each attempt was categorized as \textit{Normal, Crash}, or \textit{Glitch}.}
\change{If the fault observer} returned the expected value, the result was categorized as \textit{Normal}. If the response from the target was all 0x00 or 0xFF, then the result was categorized as \textit{Crash}, since at this point the target chip cannot behave normally unless a hard reset is performed. If the result \change{from} the fault observer was different from the expected value, but the target chip still operated normally, then it was categorized as \textit{Glitch}.  The probe was positioned using a 1mm by 1mm grid, overlayed on the target chip. A glitch attempt was performed at each position for 25 iterations in total. An EM pulse with a strength of 200V and length of 100ns, using a clockwise winded probe with a 1mm diameter, was used for profiling. A heatmap was created of each categorized result (Figure \ref{fig:heatmap}).

As shown in Figure \ref{fig:heatmap}, it is clear that Target 1 is susceptible to the glitching attack at multiple locations. Whereas Target 2 and 3 can be only glitched or crashed if the EM pulse is sent around the perimeter of the chip. Nevertheless, through profiling, it is clear that a fault can be injected using EM pulses, to affect the code running on the eMMC controller for each target.
The heatmap was then compared  with the internal structure of each target, using X-rays, and the location of the flash controller in each target chip was identified. Figure \ref{fig:Target-xray} shows the X-ray image of each target. Based on the bonding wires of the internal controller, the estimated location of the controller is highlighted with an arbitrary color.

For Target 1, the location of the ``hot spot'', where the fault observer can be successfully glitched, matches the location of the controller (Figure \ref{fig:heatmap-MAG2} and \ref{fig:xray1}). Therefore it is possible to assume that the fault is directly injected to the logic of the controller. On the other hand, the hot spot of Target 2 does not match the location of the controller (Figure \ref{fig:heatmap-JET} and \ref{fig:xray2}). Rather, crashes or glitches appear to happen when the EM pulse is sent directly on top of one of the bonding wires. This bonding wire is, to our best knowledge, related to VCC or GND. 
It is also worth mentioning that the controller of Target 2 is located \emph{under} the flash memory dies. Therefore the distance between the EM probe and the controller of this target is larger than Target 1 or Target 3. The fault observing procedure was repeated with EM pulses at a higher voltage, however the result were the same as shown in Figure \ref{fig:heatmap-JET}.  The hot spot of Target 3 matches the location of the controller, however the glitching rate is much lower (less than 10 \%), \change{compared to} Target 1 (around 30 \%), \change{for the best location}.  

Subsequently, \change{the optimal glitching parameters for each target chip were determined}. Different voltages and lengths of the EM pulse were tried at the most \change{susceptible} location of each chip (Location x=6, y=4 for Target 1, Location x=10, y=0 for Target 2, and Location x=10, y=1 for Target 3, as shown in Figure\ref{fig:heatmap}).   
\change{The pulse voltages and lengths were selected between 150V and 250V, and between 40ns and 1000ns, respectively. The parameters were randomly selected while repeating the operation for 1500 times. We observed that Target 2 and 3 are more susceptible to crashing if the voltage exceeded 200V. However, the length of the pulse did not seem to affect the result. The results were uniformly distributed regardless of the length of the pulse. Therefore, we chose the glitching parameters as shown in Table \ref{table:parameters}. The X, Y, Z position values are based on our XYZ table setup.     
}

\begin{table}[t]
\caption{Glitching parameters for each target}
\vspace{-1mm}
\centering
\begin{tabular}{|c|c|c|c|}
    \hline
    Target  & 1 & 2 & 3 \\
    \hline\hline
    X position (mm) & 154 & 159.8 & 159.8 \\
    \hline
    Y position (mm) & -62.5 & -58.1 & -58.5 \\
    \hline
    Z position (mm) & -25.7  & -25.7 & -25.7 \\
    \hline
    Pulse Voltage (V) & 200  & 200 & 200 \\
    \hline
    Pulse Duration (ns) & 100 & 100 & 100 \\
    \hline

\end{tabular}
\label{table:parameters}
\end{table}

\subsection{Firmware Reverse Engineering}
\label{sec:fWRW}

In order to gain better insight into the RPMB authentication implementation, we reverse engineered the firmware from Target 1. 
All available memory areas were dumped from the chip, including the boot ROM and main ROM code, using the aforementioned vendor commands.

The address of the reset handler was the start of our reverse engineering effort, then the location of the command handler loop was determined. The firmware uses a vector table located in SRAM, or RAM segment in case of Target 2, for all standard eMMC commands. 

\subsubsection*{Fault Modeling}
Commands CMD24 (WRITE\_BLOCK) and CMD25 (WRITE\_MULTIPLE\_BLOCK) are handled by the same function and implements all RPMB functionality. This function was further analyzed in order to understand how RPMB key authentication could be circumvented using fault injection. 

Listing \ref{src:hmac-routine} is the decompiler output for the routine that checks the HMAC of an RPMB write request. 
\begin{code}
\begin{lstlisting}[style=CStyle, caption=Routine that checks the RPMB HMAC, label=src:hmac-routine]
uint32_t rpmb_check_hmac(void *hmac,uint32_t length) {
  uint32_t i = 0;
  
  if (length + 3 >> 2 != 0) {
    do {
      if (*(int *)((int)hmac + i * 4) != *(int *)(CORRECT_HMAC + i * 4 + 0x60)) {
        return 0;
      }
      i = i + 1;
    } while (i < length + 3 >> 2);
  }
  return 1;
}

\end{lstlisting}
\end{code}
\vspace{5pt}%
The routine checks the HMAC from the eMMC packet against a pre-calculated HMAC stored in a hardware register, four bytes at a time. It returns 1 if the HMAC is valid, otherwise it returns 0. This routine was also implemented in Target 3. The following fault injection possibilities were observed:

\begin{enumerate}
    \item If the length argument is set to 0 the check is skipped in its entirety     
    \item If the register r0 is set to any non-zero value, the ROM assumes the HMAC was valid
    \item If we skip the call to \textit{rpmb\_check\_hmac} entirely, the verification will succeed, because r0 contains a pointer to the provided HMAC (non-zero value)
\end{enumerate}
\section{Glitching RPMB Authentication}
\label{section:experiment}
\subsection{Glitching Setup}
\label{section:glitching_setup}
\begin{figure}[ht!]
\begin{subfigure}{\linewidth}
\includegraphics[width = \linewidth, trim = {0, 350, 0, 0}]{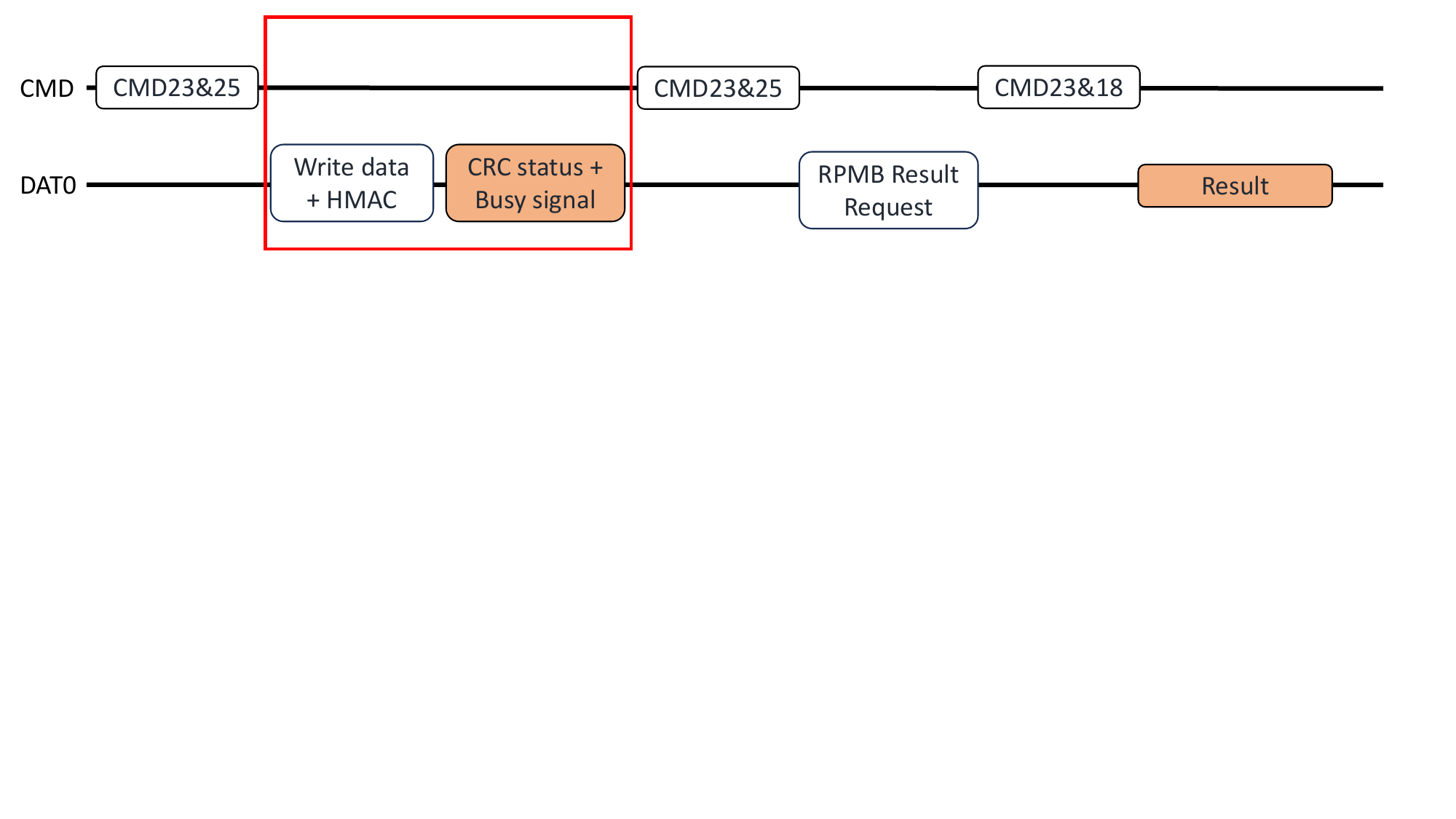}
\vspace{-12pt}
\caption{RPMB write command sequence}
\label{fig:rpmb-cmd}
\end{subfigure}
\begin{subfigure}{\linewidth}
\includegraphics[width = \linewidth, trim = {0, 0, 80, 0}]{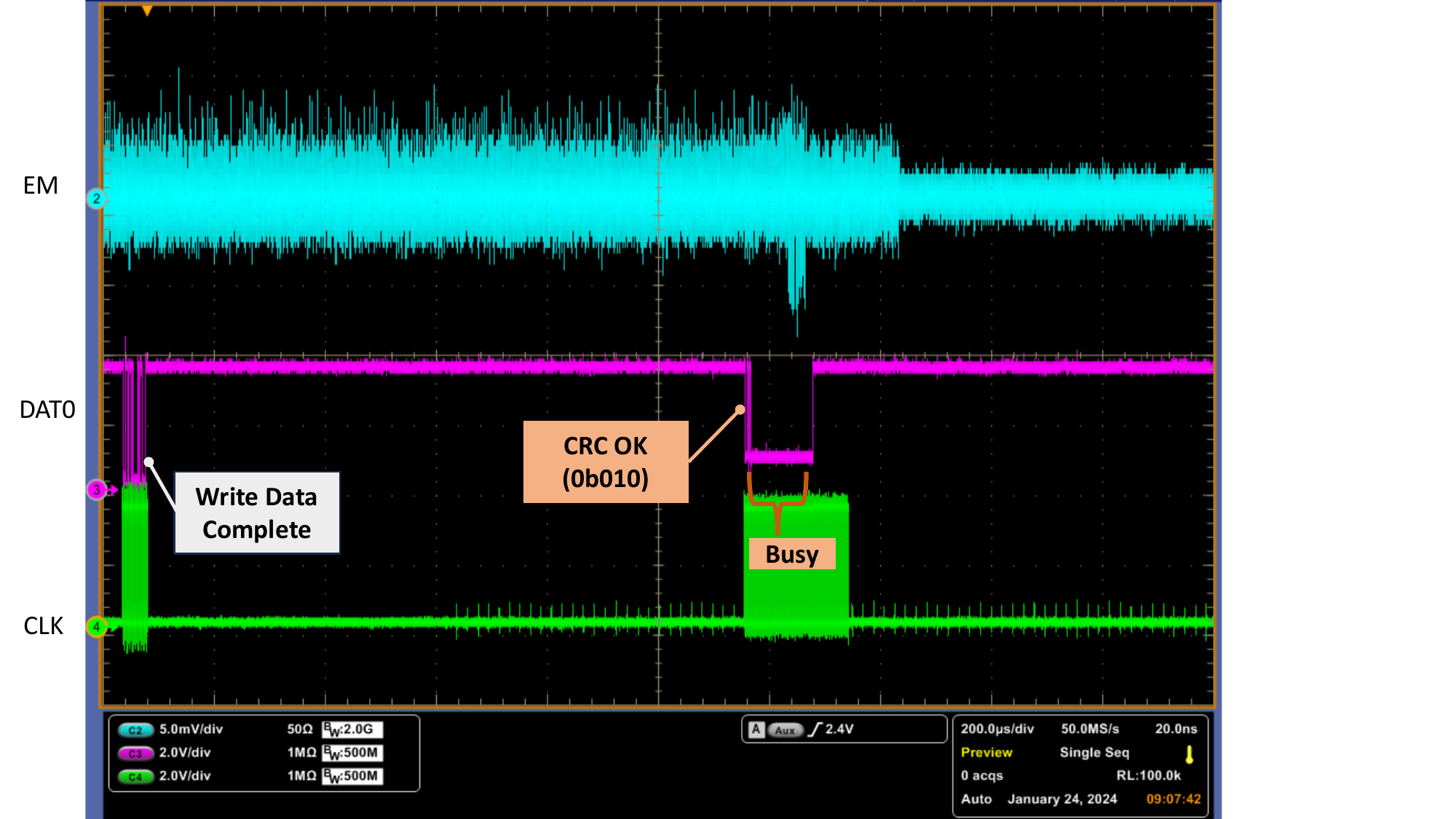}
\caption{Waveforms when data with wrong HMAC were sent to the target. The timing matches the timing window shown in red in Figure \ref{fig:rpmb-cmd}}
\label{fig:em1}
\end{subfigure}
\vspace{-10pt}
\caption{RPMB write command scheme and captured EM emissions}
\label{fig:RPMB:EM radiation}
\end{figure}

Based on the results presented in Section \ref{section:profiling}, we hypothesized that \change{it should be possible to skip the RPMB HMAC authentication routine using EMFI}.
When writing data to the RPMB, a JEDEC-defined data packet needs to be sent to the target chip. The data packet \change{is 512-bytes long, and should} contain the data to be written, Nonce, the write counter, address, block count, request message, and an HMAC calculated over this data \change{\citep{eMMC5ElecStandard}}. Upon receiving the \change{data packet}, the controller in the eMMC calculates the HMAC using the previously programmed key. If the HMAC matches the received one, the controller starts the data writing operation. The command sequence of the RPMB write routine is shown in Figure \ref{fig:rpmb-cmd}.  Commands and data in white boxes are sent from the host to the target. Orange boxes indicate responses from the target. RPMB authentication is most likely performed during the time indicated by the red box, thus the timing for performing the fault injecting attack is critical.

In order to precisely identify the attack timing window, the EM radiation emitted from the controller was measured at the moment when the RPMB data packet was sent to the target. Figure \ref{fig:em1} shows the captured waveform. The captured timing matches the timing window indicated by the red box in Figure \ref{fig:rpmb-cmd}.
DAT0 line is shown in magenta, CLK line in green, and EM emission in blue. As shown in the figure, during and after the data is sent to the target, the controller keeps performing internal operations. The CRC status of the sent data (positive = ``010'') is sent on the DAT0 line from the controller, synchronized with the CLK signal, \change{as defined by JEDEC Standard \cite{eMMC5ElecStandard}}. This operation is followed by DAT0 line pulled low, which signals that the target device is busy performing the internal computations. Even after the DAT0 goes back to high, the operation continues on the controller. \change{Since no other commands are issued to the target, we made an assumption that the HMAC verification is most likely performed during this timeframe, and that the observed EM emission comes from the controller processing the RPMB data and performing the HMAC computation.}

After sending the RPMB result request, the host can request reading the result register value, which is initiated by a read command  (CMD23 (SET\_BLOCK\_COUNT) and CMD18(READ\_MULTIPLE\_\\BLOCK), shown in Figure \ref{fig:rpmb-cmd}). Table \ref{table:rpmbresults} shows the
list of the result register values defined by JEDEC \citep{eMMC5ElecStandard}. Whilst more values are defined, only the relevant values in this setup are listed in Table \ref{table:rpmbresults}.   

\begin{table}[h!]
\caption{Partial list of RPMB operation result register values}
\label{table:rpmbresults}
\centering
\begin{tabular}{|c|c|}
    \hline
    Value & Results \\
    \hline
    0x00 & Operation OK\\
    \hline
    0x01 & General failure (Multiple errors have occurred)\\
    \hline
    0x02 & Authentication failure (HMAC mismatch) \\
    \hline
    0x03 & Counter failure \\
    \hline
    0x04 & Address failure \\
    \hline
    \end{tabular}
\end{table}

The returned register values were monitored to determine if the fault was successfully injected into the RPMB authentication procedure. If the returned value was 0x02, the target was determined to be responding normally. Because we intentionally sent an incorrect HMAC value, this is the expected result.
If the target responded with value 0x00, the attack had successfully skipped the authentication procedure. If any other value was returned, then it was determined that an error occurred during the RPMB authentication procedure.
If all of the responses were 0x00 or 0xff, the target was presumed to have crashed, since a hard reset is required to bring the chip back to a normal working state.

The setup described in Section \ref{sec:glitchingsetup} was reused, however the Raspberry Pi Pico was reprogrammed to communicate with the RPMB on the target eMMC. The JEDEC eMMC Electrical Standard \citep{eMMC5ElecStandard} was followed to send the required command sequence. First, the RPMB was programmed with an arbitrary key.  Then the first block of the RPMB (256 bytes) was programmed with random values. Subsequently the RPMB write routine was repeated with the wrong HMAC value and the EM pulse was introduced during the above mentioned timing window. After completing the write operation and waiting for the busy signal to be cleared, the RPMB result request was sent to the target, followed by the result reading command. The 200V EM pulse with a length of 100ns  was injected at 10ns granularity during the target timing window. \change{The trigger signal was generated when the last bit of the data packet was sent. The time-frame between the trigger and the end of the busy signal is around 119us for Target 1, and 113us for Target 3.} The procedure was repeated until the target responded with the "Operation OK" register value, and the correct response. \change{Additionally, in case the system data got corrupted, vendor-specific commands were implemented in the PIO state machine of the Raspberry Pi Pico, which enabled us to restore the system data.}

\subsection{Results}

\begin{figure*}
\centering
\begin{subfigure}{\linewidth}
\centering
\includegraphics[width = 0.8\linewidth, trim = {0, 190, 150, 0}]{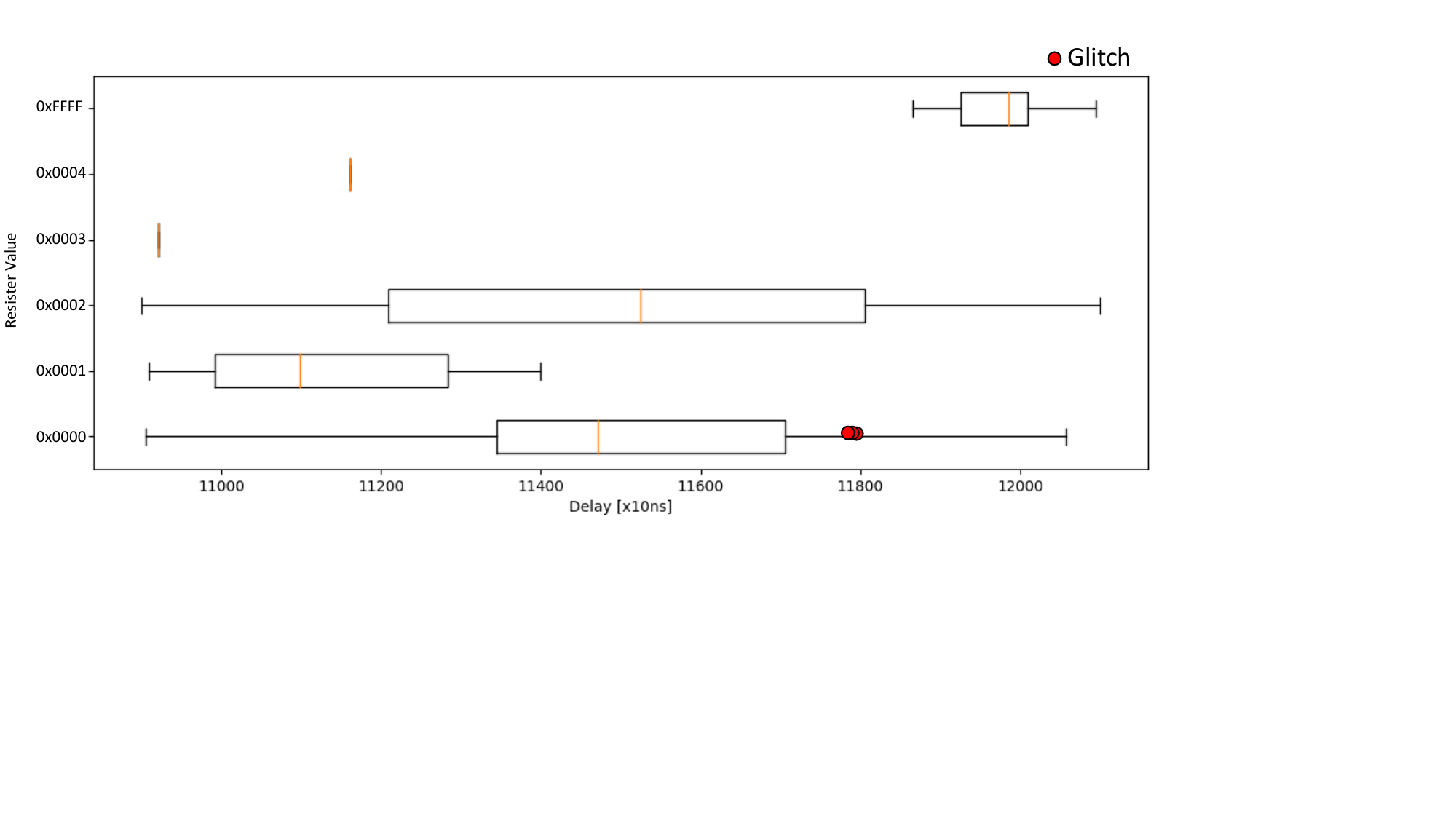}
\caption{Target 1}
\label{fig:rpmbglitchresultT1}
\end{subfigure}
\begin{subfigure}{\linewidth}
\centering
\includegraphics[width = 0.8\linewidth, trim = {0, 190, 150, 0}]{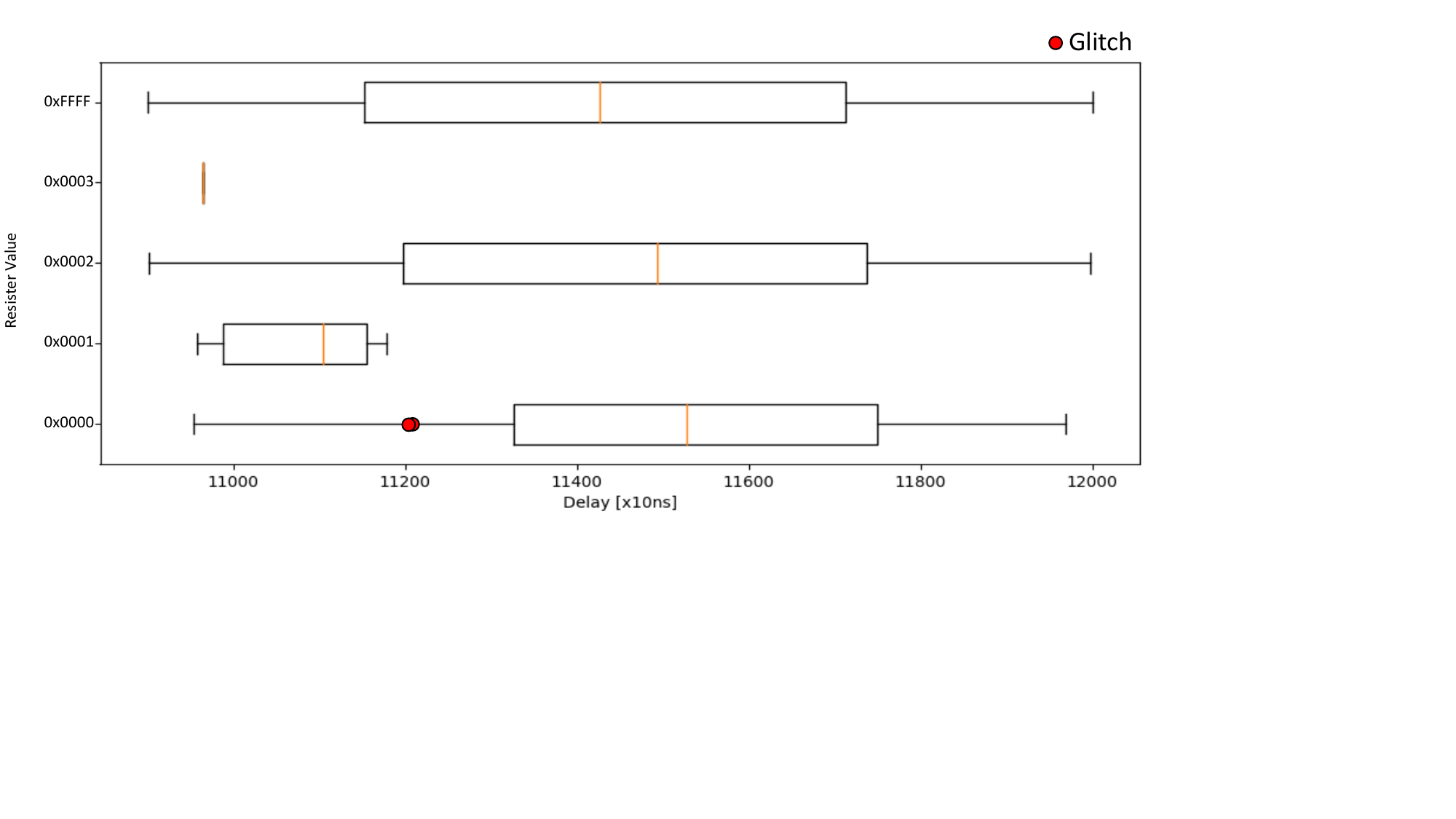}
\caption{Target 3}
\label{fig:rpmbglitchresult2}
\end{subfigure}
\caption{Returned result register value after RPMB authentication glitching}
\label{fig:rpmbglitchresult}
\end{figure*}

The glitching campaign targeting RPMB authentication was executed exclusively on Target 1 and 3. During the profiling campaign, Target 2 became non-responsive, resulting in the unavailability of samples for analysis.
Figure \ref{fig:rpmbglitchresultT1}  shows the returned result register values over time on Target 1. The x-axis shows the timing where the EM pulse was injected, and the y-axis shows the actual returned value of the result register. The timing is shown as a delay since the last bit of write data was sent to the target. Responses with 0xFFFF and 0x0000 occur mostly when the target has crashed. Register values 0x01, 0x03 and 0x04 are returned multiple times by introducing the EM pulse.  
The assumption is that the internal operations were corrupted through EMFI, while the rest of the RPMB routine continued executing on the controller.
Additionally, the write counter value was also \change{overwritten with an unexpected value several times} 
on Target 1, because it is stored in SRAM. \change{We had to restore this value by using vendor-specific commands before continuing the glitching campaign.}
Nevertheless, the RPMB authentication was successfully bypassed at the timing shown as red dots in Figure \ref{fig:rpmbglitchresultT1}. 

After successfully glitching Target 1, we repeated the same procedure on Target 3 by \change{using the parameters defined in Table \ref{table:parameters}.}
Figure \ref{fig:rpmbglitchresult2} shows the returned result register value. Similar to what was observed on Target 1, 0x01, or ``General failure'' was often observed at the early stage of the RPMB authentication. At the same time, Target 3 crashed more often than Target 1, where the response packet was filled with 0xFFFF. Nevertheless, the RPMB authentication was successfully bypassed when the EM pulse was injected around 112us, as shown in Figure \ref{fig:rpmbglitchresult2}. Unlike Target 1, no data corruption was observed during the glitching campaign on Target 3.

\change{The critical timing moment is from 117.72us to 118.30us for Target 1, and from 112.3us to 112.50us for Target 3 since the trigger (after the last bit of the data packet was sent from the PIO)}, where allegedly one of the scenarios suggested in Section \ref{sec:fWRW} was executed. On both targets, those timings are around the end of the busy signal indicated on the DAT0 line. After successful glitching, the counter value was incremented by one on both targets, which means that the controller acted as if the correct HMAC was received and proceeded with writing the received value. Post reading of the RPMB value confirmed that the sent data was successfully stored. Therefore we concluded that the RPMB authentication can be successfully skipped through EMFI if it is performed at the correct timing.



\subsection{Integrity of Non-Volatile Data}
\begin{figure}[!ht]
\centering
\includegraphics[width = .8\linewidth, trim = {0, 400, 900, 450}, clip]{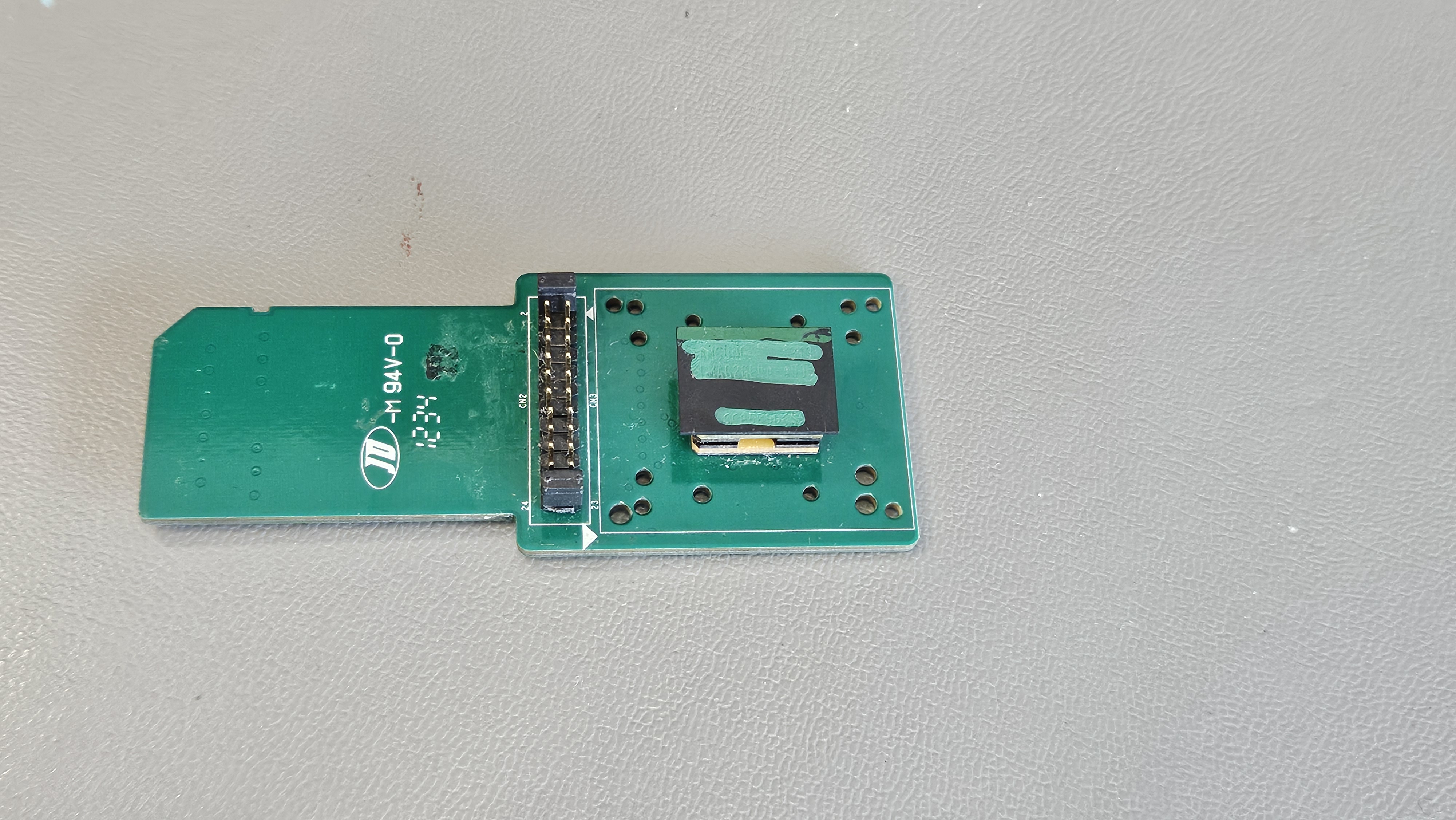}%
\caption{Target device mounted on an SD-eMMC Adaptor}
\label{fig:sdadoptor}
\end{figure}

Since the critical timing to skip the RPMB authentication check is identified, we repeated the experiment with a new chip. Prior to the experiment, arbitrary data is written into the user data area. Before starting the glitching campaign, all physical data was dumped and hashed using SHA-256. Figure \ref{fig:sdadoptor} shows the mounting of an eMMC on a custom-made eMMC-SD adapter. The adapter was connected to a PC using an SD-card slot. The data is imaged using the \texttt{dd} command on a Linux operating system (OS). 

\change{After completing the non-volatile data extraction, the target chip was mounted on the glitching setup, and the EM pulse was injected only at the critical timing of the RPMB authentication, identified above. 
Both on Target 1 and 3, bypassing the RPMB authentication was successful in less than 10 tries. After successfully bypassing the RPMB authentication, the user data area was again extracted. The SHA-256 hash value of the extracted data successfully matched the value computed before the glitching attack. Additionally, the RPMB data was only overwritten at the specified block address, keeping the remaining block data unchanged. Therefore we conclude that EMFI attacks against RPMB authentication can be conducted, while preserving integrity of the stored data, by introducing pulses using predetermined parameters, including the timing.}

It must be noted, however, that repeatedly sending EM pulses to the same device does seem to increase the chance of data corruption in the user data area of the eMMC. We repeated the glitching campaign on Target 1 and Target 3 for additional 100 consecutive times using our setup. We observed multiple data corruptions in the user data area of both targets, while the RPMB area did not seem to be affected. Any corruptions could easily be reverted using the write command to the affected data sectors. However this observation emphasizes the importance of creating a data backup beforehand, and checking the integrity of the data after a successful glitching attempt. 

\section{Discussion}
\label{section:discussion}
\subsection{Arbitrary Code Execution}
It is possible to run our own fault observer code on the target during the characterization phase due to vendor-specific commands provided by the manufacturer. Although convenient, this is not a requirement for device characterization.
The response from the eMMC controller, after requesting the result for an RPMB write request, includes a result register \cite{eMMC5ElecStandard}, as noted in Table \ref{table:rpmbresults} in Section \ref{section:glitching_setup}. This register indicates which operation failed while processing the request (e.g. HMAC mismatch, expired write counter, etc.). It was observed that applying EMFI to the target while it processes an RPMB write request, changes this value. Therefore other methods can be used for characterization, that do not rely on arbitrary code execution.

\subsection{Timing}
Our initial setup relied on the Linux kernel eMMC driver for communication with the target device. However, the scheduling properties of a non real-time operating system, makes it unsuitable for FI purposes. 
Using the Linux kernel driver also lacks the required fine-grained control needed for communication. The kernel driver periodically sends commands, that are not controlled by the user, to check the status of the device. Therefore programmable input/output (PIO) on the Raspberry Pi Pico was used instead.
A benefit of this approach is that the timing can be precisely controlled. However, this also means that a deep understanding of the eMMC protocol is required to correctly implement the PIO state machines, for communication with the target device.

Although various open source implementations are available \cite{sdio_zuluscsi,sdio_pico}, they eventually did not meet the requirements for this experiment, and therefore a custom implementation was used.

\subsection{Data Corruption}
\change{By sending high-voltage EM pulses to the target device, we observed both user data and system data corruption. Especially through profiling, where EM pulses are shot directly on the flash memory dies, some sectors became unresponsive. In some cases, the target eMMC became completely unresponsive, preventing further analysis. These results show that repeating EM pulses are highly destructive on eMMCs. Therefore it is important to perform the profiling using a reference device to identify the correct timing and location. Furthermore, creating a backup of the stored data before executing the EMFI attack is highly recommended.}

\subsection{Real World Applications}

Even though use of the RPMB, to the best of our knowledge, has only seen limited applications in smartphones (e.g. storing an anti-rollback counter \cite{DFRWS2024}), use of this type of storage seems to be embraced by the automotive industry. For example, U-Boot is a popular bootloader used in embedded devices, and stores anti-rollback counters and the bootloader lock state in the RPMB \cite{uboot_rpmb}, when Android Verified Boot (AVB) is configured to use OP-TEE. 

According to NXP, their I.MX family of processors is used by most large car manufacturers \cite{nxp_presentation}. 
NXP also supports Android Automotive, that uses Trusty as the operating system of choice to run in the TEE \cite{NXP_automotive}. According to the \textit{i.MX Android Security User's Guide} from NXP \cite{nxp_imx_security_guide}, the RPMB is used to store anti-rollback counters, bootloader lock state and AVB public key, which is used for verifying integrity of system images. The RPMB key itself is encrypted and decrypted in the TEE by Trusty. The Digi ConnectCore 8X system-on-module (SOM), which is designed around the NXP I.MX 8X processor, and uses eMMC storage, also stores the AVB public key in the RPMB \cite{ConnectCore}. 

It therefore seems that the RPMB is used as a critical component in the secure boot implementation of a wide variety of automotive products. Compromising the RPMB means that an attacker will be able to rollback potentially vulnerable versions of software, unlock the bootloader or re-sign system images. Resulting in the attacker gaining root privileges in the Android operating system.

Whilst an eMMC chip does provide tamper resistant storage in the form of an RPMB, any mitigations against fault injection have not been encountered during our experiments. Furthermore, the total cost of re-creating our FI setup is less than USD\$7.000, meaning that applying EMFI is in reach for most attackers.

\subsection{Mitigations}
As discussed in Section \ref{section:characterize}, publicly known vendor-specific commands were used to achieve code execution on all devices, and run our fault observer routine. In the case of Target 1 and 3, this already breaks the security of the RPMB, since the firmware can be patched. The firmware of Target 2 was not fully reversed.

Applying EMFI on an eMMC chip requires physical access to the device. Depending on the threat model being used, this might be a valid concern. One way to increase the difficulty of successful FI attacks is to implement mitigations in software as described by van Woudenberg and O'Flynn \cite{hardware_hacking} (e.g. double checking critical data, using non-trivial constants). 
\change{As mentioned in Section \ref{sec:fWRW}, the implemented HMAC validation routine shown in Listing \ref{src:hmac-routine} only fails when returning 0. This requirement is trivial to achieve since the CPU register holding the return value (r0) holds a pointer to the HMAC, and thus is non-zero before the function is called.}
\change{Requiring a return value with a large Hamming distance (i.e. 0xA5C3B4D2) significantly increases the attack complexity. A large number of bit flips is needed to end up with the correct return value.} 

\change{Validation of the HMAC is critical, if circumvented the integrity of the RPMB is fully compromised. Therefore the HMAC should be checked multiple times. Preferably a random delay should be added between both checks. This requires the attacker to insert multiple glitches with a non-constant delay in between.} 

\change{The implemented routine is also vulnerable to a timing attack. The function returns as soon as the comparison fails. A possible mitigation would be to always check the entire HMAC, ensuring that the comparison uses constant time.}

\change{Hardware mitigations against fault injections attack could also be considered. For example, \citeauthor{jiang2022machine} showed, using a simulated model, how machine learning can be applied to detect voltage glitching attacks in low power circuits \cite{jiang2022machine}. \citeauthor{ruminot2023novel} proposed adding a potentiometer, with a random resistance whenever the device is started, coupled with a capacitor, in order to mitigate the effect of voltage glitching attacks \cite{ruminot2023novel}. The authors suggest integrating this circuit within the same package as the microcontroller.}

\subsection{Future Work}
This research focused on attacking the RPMB embedded in eMMC devices. However succeeding technologies, such as UFS and NVMe, also include an RPMB implementation. UFS is the storage solution of choice for the current generation of smartphones making it an interesting subject for future research. 




\change{As described in Section \ref{section:threat_model}, by compromising the integrity of the RPMB data, an attacker could take over the target system. Expanding this research into a consumer device, such as automotive devices, represents a promising avenue for further exploration.}


\section{Responsible Disclosure}
We have reported our findings to Samsung Device Solution Product Security Incident Response Team. They thoroughly investigated the issue and assigned CVE-2024-31955 to the identified vulnerability.

\section{Conclusion}
\label{section:conclusion}
We have shown that it is possible to circumvent the RPMB authentication scheme and write arbitrary data, by applying EMFI on eMMCs \change{from a major manufacturer}. The RPMB functionality of non-volatile storage devices is often considered to store data that requires to be immutable to the user. \change{The potential for bypassing RPMB authentication and manipulating its data raises concerns regarding the integrity of security mechanisms, including}
anti-rollback protection, bootloader lock state, and signature verification.
Widely available, off-the-shelf commercial components were used for these experiments. 
The total cost of the equipment, needed to re-create the setup, is less than USD\$7.000, which makes it a feasible option for most attackers.



%

   \bibliographystyle{IEEEtranN}
   \bibliography{biblist}

\begin{thebibliography}{36}
\providecommand{\natexlab}[1]{#1}
\providecommand{\url}[1]{#1}
\csname url@samestyle\endcsname
\providecommand{\newblock}{\relax}
\providecommand{\bibinfo}[2]{#2}
\providecommand{\BIBentrySTDinterwordspacing}{\spaceskip=0pt\relax}
\providecommand{\BIBentryALTinterwordstretchfactor}{4}
\providecommand{\BIBentryALTinterwordspacing}{\spaceskip=\fontdimen2\font plus
\BIBentryALTinterwordstretchfactor\fontdimen3\font minus \fontdimen4\font\relax}
\providecommand{\BIBforeignlanguage}[2]{{%
\expandafter\ifx\csname l@#1\endcsname\relax
\typeout{** WARNING: IEEEtranN.bst: No hyphenation pattern has been}%
\typeout{** loaded for the language `#1'. Using the pattern for}%
\typeout{** the default language instead.}%
\else
\language=\csname l@#1\endcsname
\fi
#2}}
\providecommand{\BIBdecl}{\relax}
\BIBdecl

\bibitem[Cai(2019)]{qc_rpmb}
\BIBentryALTinterwordspacing
L.~Cai, ``Guard your data with the qualcomm snapdragon mobile platform,'' apr 2019, [Online; accessed 4. Feb. 2024]. [Online]. Available: \url{https://www.qualcomm.com/content/dam/qcomm-martech/dm-assets/documents/guard_your_data_with_the_qualcomm_snapdragon_mobile_platform2.pdf}
\BIBentrySTDinterwordspacing

\bibitem[Fukami et~al.(2024)Fukami, Buurke, and Geradts]{DFRWS2024}
A.~Fukami, R.~Buurke, and Z.~Geradts, ``{Exploiting RPMB authentication in a closed source TEE implementation},'' \emph{Forensic Science International: Digital Investigation}, vol.~48, 2024.

\bibitem[Giese and Noubir(2021)]{giese2021amazon}
D.~Giese and G.~Noubir, ``Amazon echo dot or the reverberating secrets of iot devices,'' in \emph{Proceedings of the 14th ACM Conference on Security and Privacy in Wireless and Mobile Networks}, 2021, pp. 13--24.

\bibitem[{Digi International Inc.}(2024)]{ConnectCore}
\BIBentryALTinterwordspacing
{Digi International Inc.}, ``{Secure boot flow {$\vert$} ConnectCore 8X},'' Feb. 2024, [Online; accessed 2. Feb. 2024]. [Online]. Available: \url{https://www.digi.com/resources/documentation/digidocs/embedded/android/dea11/cc8x/android-trustfence_r_secure-boot-flow}
\BIBentrySTDinterwordspacing

\bibitem[{The U-Boot development community}(2021)]{uboot_rpmb}
\BIBentryALTinterwordspacing
{The U-Boot development community}, ``{Android Verified Boot 2.0 {\ifmmode---\else\textemdash\fi} Das U-Boot unknown version documentation},'' Apr. 2021, [Online; accessed 4. Feb. 2024]. [Online]. Available: \url{https://docs.u-boot.org/en/v2021.04/android/avb2.html?highlight=rpmb#avb-using-op-tee-optional}
\BIBentrySTDinterwordspacing

\bibitem[NXP(2024)]{NXP_automotive}
\BIBentryALTinterwordspacing
NXP, ``{Android Automotive for i.MX Applications Processors},'' Jan. 2024, [Online; accessed 5. Feb. 2024]. [Online]. Available: \url{https://www.nxp.com/design/design-center/software/embedded-software/i-mx-software/android-automotive-os-for-i-mx-applications-processors:ANDROID-AUTO}
\BIBentrySTDinterwordspacing

\bibitem[Tang et~al.(2017)Tang, Sethumadhavan, and Stolfo]{clkscrew}
\BIBentryALTinterwordspacing
A.~Tang, S.~Sethumadhavan, and S.~Stolfo, ``{CLKSCREW}: Exposing the perils of {Security-Oblivious} energy management,'' in \emph{26th USENIX Security Symposium (USENIX Security 17)}.\hskip 1em plus 0.5em minus 0.4em\relax Vancouver, BC: USENIX Association, Aug. 2017, pp. 1057--1074. [Online]. Available: \url{https://www.usenix.org/conference/usenixsecurity17/technical-sessions/presentation/tang}
\BIBentrySTDinterwordspacing

\bibitem[Kim et~al.(2014)Kim, Daly, Kim, Fallin, Lee, Lee, Wilkerson, Lai, and Mutlu]{rowhammer}
Y.~Kim, R.~Daly, J.~Kim, C.~Fallin, J.~H. Lee, D.~Lee, C.~Wilkerson, K.~Lai, and O.~Mutlu, ``Flipping bits in memory without accessing them: An experimental study of dram disturbance errors,'' \emph{ACM SIGARCH Computer Architecture News}, vol.~42, no.~3, pp. 361--372, 2014.

\bibitem[Shepherd et~al.(2021)Shepherd, Markantonakis, {van Heijningen}, Aboulkassimi, Gaine, Heckmann, and Naccache]{SHEPHERD2021102471}
\BIBentryALTinterwordspacing
C.~Shepherd, K.~Markantonakis, N.~{van Heijningen}, D.~Aboulkassimi, C.~Gaine, T.~Heckmann, and D.~Naccache, ``Physical fault injection and side-channel attacks on mobile devices: A comprehensive analysis,'' \emph{Computers \& Security}, vol. 111, p. 102471, 2021. [Online]. Available: \url{https://www.sciencedirect.com/science/article/pii/S0167404821002959}
\BIBentrySTDinterwordspacing

\bibitem[Sa{\ss} et~al.(2023)Sa{\ss}, Mitev, Sadeghi, and VFI]{sass2023oops}
X.~M. Sa{\ss}, R.~Mitev, A.-R. Sadeghi, and V.~F.~I. VFI, ``{Oops..! I Glitched It Again! How to Multi-Glitch the Glitching-Protections on ARM TrustZone-M},'' \emph{arXiv preprint arXiv:2302.06932}, 2023.

\bibitem[Qiu et~al.(2019)Qiu, Wang, Lyu, and Qu]{Pengfei2019}
\BIBentryALTinterwordspacing
P.~Qiu, D.~Wang, Y.~Lyu, and G.~Qu, ``Voltjockey: Breaching trustzone by software-controlled voltage manipulation over multi-core frequencies,'' in \emph{Proceedings of the 2019 ACM SIGSAC Conference on Computer and Communications Security}, ser. CCS '19.\hskip 1em plus 0.5em minus 0.4em\relax New York, NY, USA: Association for Computing Machinery, 2019, p. 195–209. [Online]. Available: \url{https://doi.org/10.1145/3319535.3354201}
\BIBentrySTDinterwordspacing

\bibitem[B.V.(2020)]{nxp_imx_security_guide}
\BIBentryALTinterwordspacing
N.~B.V., ``{i.MX Android™ Security User's Guide},'' May 2020, [Online; accessed 5. Feb. 2024]. [Online]. Available: \url{https://community.nxp.com/pwmxy87654/attachments/pwmxy87654/imx-processors/167888/2/i.MX_Android_Security_User's_Guide.pdf}
\BIBentrySTDinterwordspacing

\bibitem[And(2024)]{AndroidRollbackProtection}
\BIBentryALTinterwordspacing
``{Verifying Boot},'' Mar. 2024, [Online; accessed 29. Mar. 2024]. [Online]. Available: \url{https://source.android.com/docs/security/features/verifiedboot/verified-boot#rollback-protection}
\BIBentrySTDinterwordspacing

\bibitem[Tru(2024)]{TrustIssues}
\BIBentryALTinterwordspacing
``{Trust Issues: Exploiting TrustZone TEEs},'' Mar. 2024, [Online; accessed 29. Mar. 2024]. [Online]. Available: \url{https://googleprojectzero.blogspot.com/2017/07/trust-issues-exploiting-trustzone-tees.html}
\BIBentrySTDinterwordspacing

\bibitem[Arm(2021{\natexlab{a}})]{ArmTFACoT}
\BIBentryALTinterwordspacing
Arm, ``{Trusted Board Boot Requirements Client (TBBR-CLIENT) Armv8-A},'' \emph{Arm}, Dec. 2021. [Online]. Available: \url{https://developer.arm.com/documentation/den0006/latest}
\BIBentrySTDinterwordspacing

\bibitem[Nakamoto(2016)]{nakamoto2016secure}
R.~P. Nakamoto, ``Secure boot and image authentication,'' \emph{Qualcomm Technologies Inc., San Diego}, 2016.

\bibitem[AVB(2024)]{AVB2}
\BIBentryALTinterwordspacing
``{Android Verified Boot 2.0},'' Apr. 2024, [Online; accessed 1. Apr. 2024]. [Online]. Available: \url{https://android.googlesource.com/platform/external/avb/+/main/README.md}
\BIBentrySTDinterwordspacing

\bibitem[Jeong et~al.(2013)Jeong, Lee, Sung, and Hong]{kitae2013}
K.~Jeong, Y.~Lee, J.~Sung, and S.~Hong, ``Security analysis of hmac/nmac by using fault injection,'' \emph{Journal of Applied Mathematics}, vol. 2013, 01 2013.

\bibitem[Belenky et~al.(2021)Belenky, Dushar, Teper, Chernyshchyk, Azriel, and Kreimer]{yaacov2021}
Y.~Belenky, I.~Dushar, V.~Teper, H.~Chernyshchyk, L.~Azriel, and Y.~Kreimer, \emph{First Full-Fledged Side Channel Attack on HMAC-SHA-2}, 10 2021, pp. 31--52.

\bibitem[{Western Digital}(2020)]{western2}
{Western Digital}, ``Replay protected memory block (rpmb) - protocol vulnerabilities,'' White Paper, 2020.

\bibitem[Gangolli et~al.(2022)Gangolli, Mahmoud, and Azim]{electronics11132023}
\BIBentryALTinterwordspacing
A.~Gangolli, Q.~H. Mahmoud, and A.~Azim, ``A systematic review of fault injection attacks on iot systems,'' \emph{Electronics}, vol.~11, no.~13, 2022. [Online]. Available: \url{https://www.mdpi.com/2079-9292/11/13/2023}
\BIBentrySTDinterwordspacing

\bibitem[Bozzato et~al.(2019)Bozzato, Focardi, Palmarini, et~al.]{bozzato2019shaping}
C.~Bozzato, R.~Focardi, F.~Palmarini \emph{et~al.}, ``Shaping the glitch: optimizing voltage fault injection attacks,'' \emph{IACR transactions on cryptographic hardware and embedded systems}, vol. 2019, no.~2, pp. 199--224, 2019.

\bibitem[Ruminot et~al.(2023)Ruminot, Estevez, and Montejo-S{\'a}nchez]{ruminot2023novel}
N.~Ruminot, C.~Estevez, and S.~Montejo-S{\'a}nchez, ``A novel approach of a low-cost voltage fault injection method for resource-constrained iot devices: Design and analysis,'' \emph{Sensors}, vol.~23, no.~16, p. 7180, 2023.

\bibitem[Werner et~al.(2023)Werner, Maingault, and Potet]{werner2023end}
V.~Werner, L.~Maingault, and M.-L. Potet, ``An end-to-end approach to identify and exploit multi-fault injection vulnerabilities on microcontrollers,'' \emph{Journal of Cryptographic Engineering}, vol.~13, no.~2, pp. 149--165, 2023.

\bibitem[Avraham(2018)]{Avraham_2018}
\BIBentryALTinterwordspacing
O.~Avraham, ``{eMMC hacking, or: how I fixed long-dead Galaxy S3 phones },'' Jan 2018, [Online; accessed 23. Jan. 2024]. [Online]. Available: \url{https://media.ccc.de/v/34c3-8784-emmc_hacking_or_how_i_fixed_long-dead_galaxy_s3_phones}
\BIBentrySTDinterwordspacing

\bibitem[oranav(2024)]{oranav2024Jan}
\BIBentryALTinterwordspacing
oranav, ``{i9300{$\_$}emmc{$\_$}toolbox},'' Jan. 2024, [Online; accessed 23. Jan. 2024]. [Online]. Available: \url{https://github.com/oranav/i9300_emmc_toolbox}
\BIBentrySTDinterwordspacing

\bibitem[{Z3X-Team}(2024)]{easyjtag}
\BIBentryALTinterwordspacing
{Z3X-Team}, ``{EASY-JTAG PLUS ACTIVATION},'' Feb. 2024, [Online; accessed 5. Feb. 2024]. [Online]. Available: \url{https://z3x-team.com/products/easy-jtag-plus-activation/}
\BIBentrySTDinterwordspacing

\bibitem[Inc.(2020)]{newae}
\BIBentryALTinterwordspacing
N.~T. Inc., ``Chipshouter kit,'' 2020, [Online; accessed 6. Feb. 2024]. [Online]. Available: \url{https://www.newae.com/products/nae-cw520}
\BIBentrySTDinterwordspacing

\bibitem[mhei(2024)]{mhei2024Jan}
\BIBentryALTinterwordspacing
mhei, ``{mmc-utils},'' Jan. 2024, [Online; accessed 23. Jan. 2024]. [Online]. Available: \url{https://github.com/mhei/mmc-utils}
\BIBentrySTDinterwordspacing

\bibitem[Arm(2021{\natexlab{b}})]{Arm2021Dec}
\BIBentryALTinterwordspacing
Arm, ``{ARMv7-M Architecture Reference Manual},'' \emph{Arm}, Dec. 2021. [Online]. Available: \url{https://developer.arm.com/documentation/ddi0403/latest}
\BIBentrySTDinterwordspacing

\bibitem[{JEDEC Solid State Technology Association}(2015)]{eMMC5ElecStandard}
\BIBentryALTinterwordspacing
{JEDEC Solid State Technology Association}, ``Embedded multi-media card {(e$\cdot$MMC)} electrical standard {(5.1)},'' JEDEC Standard JESD84-B51, February 2015. [Online]. Available: \url{https://www.jedec.org/system/files/docs/JESD84-B51.pdf}
\BIBentrySTDinterwordspacing

\bibitem[Zuluscsi(2024)]{sdio_zuluscsi}
\BIBentryALTinterwordspacing
Zuluscsi, ``{ZuluSCSI-firmware},'' Feb. 2024, [Online; accessed 4. Feb. 2024]. [Online]. Available: \url{https://github.com/ZuluSCSI/ZuluSCSI-firmware/blob/main/lib/ZuluSCSI_platform_RP2040/sdio_RP2040.pio}
\BIBentrySTDinterwordspacing

\bibitem[democloid(2024)]{sdio_pico}
\BIBentryALTinterwordspacing
democloid, ``{pico-sdio-example},'' Feb. 2024, [Online; accessed 4. Feb. 2024]. [Online]. Available: \url{https://github.com/democloid/pico-sdio-example}
\BIBentrySTDinterwordspacing

\bibitem[Chien(2017)]{nxp_presentation}
\BIBentryALTinterwordspacing
C.~Chien, ``{I.MX in automotive},'' 2017, [Online; accessed 5. Feb. 2024]. [Online]. Available: \url{https://www.nxp.com/docs/en/supporting-information/BL-Micro-i.MX-in-Automotive-Carl-Chien.pdf}
\BIBentrySTDinterwordspacing

\bibitem[van Woudenberg and O'Flynn(2022)]{hardware_hacking}
J.~van Woudenberg and C.~O'Flynn, \emph{The Hardware Hacking Handbook}, 1st~ed.\hskip 1em plus 0.5em minus 0.4em\relax San Francisco, CA: No starch press, 2022, ch. Chapter 14: Think of the Children: Countermeasures, Certifications and Goodbytes.

\bibitem[Jiang et~al.(2022)]{jiang2022machine}
W.~Jiang \emph{et~al.}, ``Machine learning methods to detect voltage glitch attacks on iot/iiot infrastructures,'' \emph{Computational Intelligence and Neuroscience}, vol. 2022, 2022.

\end{thebibliography}

\end{document}